\documentclass[12pt]{article}
\usepackage{amsmath,cite,epsfig,a4,times,euscript,listings,amssymb}
\usepackage[text-hat-accent]{euler}
\usepackage{graphics}
\usepackage[usenames]{color}
\renewcommand{\baselinestretch}{1.1}
\newcommand{\myTitle}[1]{\begin{center}{\bf\Huge #1}\\[5ex]\end{center}}
\newcommand{\myAuthor}[1]{\begin{center}{\Large #1}\\[2ex]\end{center}}
\newcommand{\myAffiliation}[1]{\\[1ex]{\it\large #1}}

\newcommand{\myDate}{\begin{center}{\large\today}\\[5ex]\end{center}}
\newcommand{\myAbstract}[1]{\begin{center}\renewcommand{\baselinestretch}{1}{\bf Abstract}\\[2ex]\parbox{0.8\linewidth}{\small\hspace{15pt} #1}\end{center}\vspace{\baselineskip}}

\newcommand{\myPreprint}[1]{}
\newcommand{\myKeywords}[1]{}

\newcommand{\myScript}[1]{\EuScript{#1}}

\newcommand{\Appendix}[1]{Appendix~\ref{#1}}

\newcommand{\Figure}[1]{Fig.~\ref{#1}}
\newcommand{\Equation}[1]{Eq.~(\ref{#1})}

\newcommand{\cf}{{\it cf.}}
\newcommand{\eg}{{\it e.g.}}
\newcommand{\KaTie}{{\sc Ka\hspace{-1pt}Tie}}
\newcommand{\Ord}{\myScript{O}}
\newcommand{\srac}[2]{{\textstyle\frac{#1}{#2}}}

\newcommand{\imag}{\mathrm{i}}

\newcommand{\vep}{{\bar{\epsilon}}}
\newcommand{\vepv}{\epsilon}
\newcommand{\piep}{\pi_{\epsilon}}

\newcommand{\cNLO}{a_\epsilon}

\newcommand{\lop}[2]{#1\!\cdot\!#2}
\newcommand{\Nc}{N_\mathrm{c}}

\newcommand{\sgn}{\mathrm{sgn}}
\newcommand{\Qin}{Q}
\newcommand{\dQin}{[dQ]}
\newcommand{\xP}{x}
\newcommand{\xM}{\bar{x}}

\newcommand{\zP}{z}
\newcommand{\zM}{\bar{z}}
\newcommand{\rc}{v}

\newcommand{\xiP}{\xi_0}

\newcommand{\sTot}{{S}}
\newcommand{\pP}{P}
\newcommand{\pM}{\bar{P}}

\newcommand{\alphaS}{\alpha_{\mathrm{s}}}

\newcommand{\theperp}{\scriptscriptstyle\perp}

\newcommand{\rperp}{r_{\scriptscriptstyle\perp}}

\newcommand{\kperp}{k_{\scriptscriptstyle\perp}}
\newcommand{\Kperp}{K_{\scriptscriptstyle\perp}}

\newcommand{\Born}{\mathrm{B}}
\newcommand{\Real}{\mathrm{R}}
\newcommand{\inlbl}{\chi}
\newcommand{\inbar}{\overline{\chi}}
\newcommand{\JetB}{J_{\Born}}
\newcommand{\JetR}{J_{\Real}}

\newcommand{\Atree}{\EuScript{M}}
\newcommand{\symfac}{\EuScript{S}}
\newcommand{\Mtree}[1]{\big|\EuScript{M}_{#1}\big|^2}
\newcommand{\MtreeCor}[2]{\big(\EuScript{M}\big)^2_{\mathrm{color}(#1,#2)}}
\newcommand{\MtreeCorNorm}[2]{(m)^2_{#1,#2}}
\newcommand{\MtreeSumCor}[2]{\big(\EuScript{M}\big)^2_{\mathrm{color}(#1,#2)}}
\newcommand{\MtreeHelCor}[1]{\big(\EuScript{M}\big)^2_{\mathrm{helicity}(#1)}}

\newcommand{\muF}{\mu_F}
\newcommand{\EL}[1]{\EuScript{L}#1}
\newcommand{\ELshiftSymb}{\ell}

\newcommand{\ELshiftIin}[1]{\ELshiftSymb^{\perp}_{\inlbl}\big(#1\big)}
\newcommand{\ELshiftPerp}[1]{\ELshiftSymb^{\perp}(#1)}

\newcommand{\RtimesA}[2]{\EuScript{R}^{#1}_{#2}\otimes\EuScript{A}^{#1}_{#2}}
\newcommand{\RtimesAarg}[4]{\EuScript{R}^{#1}_{#2}(#3)\otimes\EuScript{A}^{#1}_{#2}\big(#4\big)}
\newcommand{\hatRtimesA}[4]{\hat{\EuScript{R}}^{#1}_{#2}(#3)\otimes\hat{\EuScript{A}}^{#1}_{#2}\big(#4\big)}
\newcommand{\LtimesA}[2]{\EuScript{L}^{#1}_{#2}\otimes\EuScript{A}^{#1}_{#2}}
\newcommand{\LtimesAarg}[4]{\EuScript{L}^{#1}_{#2}\big(#3\big)\otimes\EuScript{A}^{#1}_{#2}\big(#4\big)}
\newcommand{\LXYZ}[2]{L^{#1}_{#2}}
\newcommand{\SXYZ}[2]{\EuScript{S}^{#1}_{#2}}
\newcommand{\soft}{\mathrm{soft}}
\newcommand{\soco}{\mathrm{soco}}
\newcommand{\Fcol}{\mathrm{F,col}}
\newcommand{\Fcolfin}{\mathrm{F,col,fin}}
\newcommand{\Fcoldiv}{\mathrm{F,col,div}}
\newcommand{\Fsoft}{\mathrm{F,soft}}
\newcommand{\Fsoftfin}{\mathrm{F,soft,fin}}
\newcommand{\Fsoftdiv}{\mathrm{F,soft,div}}
\newcommand{\Fsoftcol}{\mathrm{F,soco}}
\newcommand{\Fsoftcoldiv}{\mathrm{F,soco,div}}
\newcommand{\Fsoftcolfin}{\mathrm{F,soco,fin}}
\newcommand{\Isoft}{\mathrm{I,soft}}
\newcommand{\Isoftfin}{\mathrm{I,soft,fin}}
\newcommand{\Isoftdiv}{\mathrm{I,soft,div}}

\newcommand{\Isoftsoco}{\mathrm{I,soft/soco}}
\newcommand{\Isoftcol}{\mathrm{I,soco}}
\newcommand{\Isoftcolfin}{\mathrm{I,soco,fin}}
\newcommand{\Isoftcoldiv}{\mathrm{I,soco,div}}

\newcommand{\Icol}{\mathrm{I,col}}
\newcommand{\Icolfin}{\mathrm{I,col,fin}}
\newcommand{\Icoldiv}{\mathrm{I,col,div}}
\newcommand{\pSetN}{\{p\}_{n}}
\newcommand{\pSetNpls}{\{p\}_{n+1}}
\newcommand{\pSetNOr}{\{p\}_{n}^{\hspace{1pt}r\hspace{-4pt}/}}
\newcommand{\pSetNOri}{\{p\}_{n}^{\hspace{1pt}r\hspace{-4pt}/\hspace{0.5pt};\hspace{0.5pt}i}}
\newcommand{\pSetNOriB}{\{p\}_{n-1}^{\hspace{1pt}r\hspace{-4pt}/\hspace{0.5pt},\hspace{0.5pt}i\hspace{-3pt}/}}

\newcommand{\graph}[3]{\raisebox{-#3ex}{\epsfig{file=#1.pdf,width=#2ex}}}

\newcommand{\tweakcodepar}[3]%
  {\vspace{#1ex}\newline\noindent\hspace*{4.0ex}{\small\tt #3}\vspace{#2ex}\newline\noindent}

\begin{document}
%


\myTitle{%
A new subtraction scheme at NLO\\[-0.5ex] exploiting the privilege of\\[0.5ex] $\boldsymbol{k_T}$-factorization
}

\myAuthor{%
Alessandro Giachino$^{ab}\footnote{Alessandro.Giachino@ific.uv.es}$,
Andreas~van~Hameren$^{a}\footnote{hameren@ifj.edu.pl}$,\\
and Grzegorz Ziarko$^{a}\footnote{grzegorz.ziarko@ifj.edu.pl}$%
\myAffiliation{%
$^a$~Institute of Nuclear Physics Polisch Academy of Sciences,\\
PL-31342 Krak\'ow, Poland
%
\\[1ex]$^b$~Departamento de Física Teórica and IFIC,\\
Centro Mixto Universidad de Valencia-CSIC\\[0.5ex]
Institutos de Investigación de Paterna, 46071 Valencia, Spain
}
}

\myDate

\myAbstract{%
We present a subtraction method for the calculation of real-radiation integrals at NLO in hybrid $k_T$-factorization.
The main difference with existing methods for collinear factorization is that we subtract the momentum recoil, occurring due to the mapping from an $(n+1)$-particle phase space to an $n$-particle phase space, from the initial-state momenta, instead of distributing it over the final-state momenta.
}

\myKeywords{QCD}

%

\newpage%
\setcounter{tocdepth}{2}
\tableofcontents

\section{Introduction}
In the calculation of real radiation contributions to scattering cross sections in perturbative quantum chromodynamics (QCD) one encounters divergent phase space integrals (\cf~\cite{Somogyi:2006cz} for an accessible introduction).
On one hand, these are preferably dealt with using dimensional regularization, in order to match the divergences from virtual contributions and have a manifest cancellation of those.
On the other hand, one would like to calculate arbitrary differential distributions for arbitrary fiducial phase space regions, and preferably even generate event files, pointing at the Monte Carlo method of numerical integration as the obvious method of choice.
One way to satisfy both requirements is the subtraction method, in which the divergent integrals are regularized by subtracting extra terms that point-wise cancel the singularities causing the divergences.
These extra terms are then constructed such that they themselves can be integrated analytically within dimensional regularization, explicitly exposing the divergences ready to be cancelled against the virtual ones.
While conceptually easy, explicit implementation faces various complications especially for increasing final-state multiplicity, and over the last decades several solutions have been devised both at next-to-leading order (NLO)%
~\cite{%
Frixione:1995ms
,Catani:1996vz
,Somogyi:2006cz
,Somogyi:2009ri
,Robens:2013wga
,Bevilacqua:2013iha
,Fox:2023bma
}
and next-to-next-to-leading order (NNLO)%
~\cite{%
Gehrmann-DeRidder:2005btv
,Somogyi:2006da
,Czakon:2010td
,Caola:2017dug
,Magnea:2018hab
}%
.

All the cited approaches were designed for collinear factorization, for which the partonic initial-state momenta are light-like.
In this work, we consider $k_T$-factorization~\cite{Collins:1991ty,Catani:1990eg}, hybrid $k_T$-factorization to be precise for which only one of the initial-state partons is space-like, and its accompanying PDF depends on transverse momentum ($k_T$) besides the longitudinal momentum fraction ($x$)%
~\cite{%
Dumitru:2005gt
,Marquet:2007vb
,Deak:2009xt
}%
.
This factorization becomes phenomenologically relevant for scattering events in which the final-state products are boosted towards one initial-state direction, implying that one of the collinear momentum fractions was much smaller than the other.
It has already been applied at NLO precision for selected processes%
~\cite{%
Nefedov:2019mrg
,Nefedov:2020ecb
,Hentschinski:2020tbi
,Celiberto:2022fgx
,Bergabo:2022zhe
,Taels:2023czt
,Altinoluk:2023hfz
}%
.
The matrix elements with space-like initial-state partons cannot be obtained from a trivial kinematic continuation of completely on-shell matrix elements, and need to be defined properly.
One approach to achieve this is the auxiliary parton method~\cite{vanHameren:2012if}, which is particularly well-suited to be applied together with efficient and easily automatable methods for the calculation of matrix elements that are successfully applied in programs for collinear factorization.
The approach is implemented as such at tree-level in the parton-level Monte Carlo generator \KaTie~\cite{vanHameren:2016kkz}.
The auxiliary parton method embeds the process with space-like partons into a process in which the space-like parton is replaced by a pair of on-shell partons.
One can extract the necessary extra Feynman rules to construct the desired space-like amplitudes directly as is done in \KaTie, or one can use a complete expression of an embedding on-shell amplitude to extract the desired space-like amplitude.
The latter method was applied in~\cite{Blanco:2020akb,Blanco:2022iai}, to calculates space-like one-loop amplitudes for selected processes. 

While most divergences appearing in collinear factorization eventually cancel at the parton-level, some divergences that can be identified as being of the collinear kind remain, and are taken care of by renormalization of the PDFs.
The fact that this works in a process-independent manner and to any order in the perturbative expansion is an essential ingredient in the predictive power of collinear factorization.
In~\cite{vanHameren:2022mtk}, a scheme was outlined to perform calculations in hybrid $k_T$-factorization with the auxiliary parton method at NLO.
It was shown that, while there are more divergences that do not cancel at the parton-level, their occurrence is process independent, opening the possibility of a renormalization procedure and perform predictions.

It is also shown in~\cite{vanHameren:2022mtk} that the process-dependent part of the real-radiation contribution to the cross section is obtained just as in collinear factorization, as the phase space integrals of the processes with one more final-state parton and a jet algorithm allowing for one final-state parton to be unresolved.
It was argued that all the divergences coming from these integrals are exactly as in collinear factorization, except for the remnant collinear divergence related to the space-like initial-state parton.
In this work, we will make this more explicit, by introducing a subtraction method to tackle the real-radiation phase space integrals, and providing the universal expressions of the divergent part of the integrated subtraction terms.
%
%
%

We follow the  work of~\cite{Somogyi:2006cz,Somogyi:2009ri} as a guideline for the organization of the subtraction terms.
The main difference is the treatment of the momentum recoil appearing in the necessary mapping of the $(n+1)$-particle phase space to a $n$-particle phase space.
Since we have $4$ independent kinematic variables in the initial state ($2$ longitudinal components, and $2$ transverse ones), we can simply subtract the recoil from those.
This has the advantage that the final-state momenta are touched in a minimal way, which is helpful in the calculation of differential distributions, possibly minimizing the so-called mis-binning problem.
As a drawback, the finite part of the integrated subtraction terms cannot be obtained completely analytically and requires the application of numerical integration.
This, however, can easily be combined with the Monte Carlo procedure for the phase space integration.

\section{Exposition of the subtraction scheme}
We consider hadron scattering with light-like hadron momenta $\pP,\pM$.
For arbitrary momentum $K$, we can then make the Sudakov decomposition
%
\begin{equation}
K^\mu = \xP_K\pP^\mu + \xM_K\pM^\mu + \Kperp
\quad,\quad
\lop{\pP}{\pM} = \srac{1}{2}S>0
\label{Eq:18}~,
\end{equation}
%
with
%
\begin{equation}
\xP_K = \frac{\lop{K}{\pM}}{\lop{\pP}{\pM}}
\quad,\quad
\xM_K = \frac{\lop{K}{\pP}}{\lop{\pP}{\pM}}
\quad,\quad
\Kperp^\mu = K^\mu - \xP_K\pP^\mu - \xM_K\pM^\mu
~.
\end{equation}
%
We use the labels $\inlbl,\inbar$ to indicate the partonic initial states, and the momenta in hybrid $k_T$-factorization are
%
\begin{equation}
k_{\inlbl}^\mu = \xP\pP^\mu + \kperp^\mu
\quad,\quad
k_{\inbar}^\mu = \xM\pM^\mu
~,
\label{Eq:454}
\end{equation}
%
that is we omit the labels for the longitudinal and transverse components of those: $\xP_{\inlbl}=\xP$ and $\xM_{\inbar}=\xM$.
Eventually, we will assume that calculations are performed in a frame for which
%
\begin{equation}
\pP^\mu = (E,0,0,E)
\quad,\quad
\pM^\mu = (\bar{E},0,0,-\bar{E})
\quad,\quad\textrm{so}\quad
\sTot = 4E\bar{E}
~.
\label{Eq:153}
\end{equation}
%
The symbol $E$ with an appropriate subscript will also be used for the energy of momenta in general.
Also, we will employ unit vectors denoted with the letter $n$.
So for a light-like momentum $K^\mu$ we may write
%
\begin{equation}
K^\mu = E_Kn_K^\mu = E_K(1,\vec{n}_K)
~.
\end{equation}
%
We will use a separate symbol for the sum of the parton-level initial-state momenta:
%
\begin{equation}
\Qin^\mu = k_{\inlbl}^\mu + k_{\inbar}^\mu = \xP\pP^\mu + \xM\pM^\mu + \kperp^\mu
~.
\end{equation}
%
Integration over the initial-state variables will be abbreviated as
%
\begin{equation}
\int\dQin = \int_0^1d\xP \int_0^1d\xM \int d^2\kperp
~.
\label{Eq:dQ}
\end{equation}
%

One thing we need to address already before we continue is our notation regarding dimensional regularization.
We write
%
\begin{equation}
\mathrm{dim} = 4+\vep
\quad,\quad
\vep=-2\vepv
~,
\end{equation}
%
and will often use $\vep$ in calculations because it can be imagined to be positive for infra-red divergent integrals.
Furthermore, we define
%
\begin{equation}
\piep = \int d^{2+\vep}r\,\delta\big(1-|r|^2\big)
      = \frac{\pi^{1+\vep/2}}{\Gamma(1+\vep/2)}
      = \frac{\pi^{1-\vepv}}{\Gamma(1-\vepv)}
~,
\end{equation}
and
\begin{equation}
\cNLO
  = \frac{\alphaS}{2\pi}\,\frac{(4\pi)^{-\vep/2}}{\Gamma(1+\vep/2)}
  = \frac{\alphaS}{2\pi}\,\frac{(4\pi)^{\vepv}}{\Gamma(1-\vepv)}
~.
\end{equation}
%
The common overall factor $4\pi\alphaS\mu^{2\vepv}$ for singular limits of matrix elements divided by the phase space factor $(2\pi)^{3+\vep}$ of the radiation can be written as 
%
\begin{equation}
\frac{4\pi\alphaS\mu^{2\vepv}}{(2\pi)^{3+\vep}} = \frac{\cNLO}{\piep\mu^{\vep}}
~.
\end{equation}
%
The denominator factor $\piep$ typically drops out in final integrated expressions.

The Born-level formula for the cross section in hybrid $k_T$-factorization is
%
\begin{equation}
\sigma_{\Born}=
\frac{1}{\symfac_n}\int\dQin
\int
d\Phi\big(\Qin;\pSetN\big)
\,\EL{\big(\Qin;\pSetN\big)}
\,\Mtree{}\big(\Qin;\pSetN\big)
\,\JetB\big(\pSetN\big)
~,
\end{equation}
%
where $d\Phi$ is the differential phase space for the final-state momenta $\pSetN$
\begin{equation}
d\Phi\big(\Qin;\pSetN\big) =
  \Bigg(\prod_{l=1}^n\frac{d^4p_l}{(2\pi)^3}\delta_+(p_l^2-m_l^2)\Bigg)
  \frac{1}{(2\pi)^4}\,\delta\bigg(\Qin-\sum_{l=1}^np_l\bigg)
~,
\end{equation}
%
and $\Mtree{}\big(\Qin;\pSetN\big)$ is the tree-level matrix element of the considered parton-level process, sum\-med over helicity- and color-degrees of freedom.
It does not include symmetry factors nor factors to average over initial-state degrees of freedom -- they are captured by $\symfac_n$.
The function $\EL{}$ includes both the collinear and the $k_T$-dependent PDF, and also the flux factor
%
\begin{equation}
\EL{\big(\Qin;\pSetN\big)}
=
\frac{F_{\inlbl}\big(\xP,\kperp,\muF(\pSetN)\big)
    \,f_{\inbar}\big(\xM,\muF(\pSetN)\big)}
     {2\xP\xM\sTot}
~,
\label{Eq:179}
\end{equation}
%
where $\muF$ is factorization scale.
Like in~\cite{vanHameren:2022mtk}, we will restrict ourselves to the situation in which the $\inlbl$ parton is a space-like gluon.
The light-like $\inbar$ parton can be gluon or quark or antiquark.
The symbol $\JetB$ denotes the jet function, which at Born-level eventually consists of a number of phase space cuts making sure the singularities of $\Mtree{}$ are avoided. 

\subsection{Limits of matrix elements}
In the following, we assume that all final-state parton momenta are light-like.
The real-radiation contribution at NLO involves scattering processes with one more final-state parton than at Born level.
Furthermore, the jet function, now denoted $\JetR$, does not avoid all singularities of the tree-level squared matrix element anymore, but allows one pair of partons to become collinear, 
%
\begin{equation}
p_r||p_i \quad\Leftrightarrow\quad \vec{n}_r-\vec{n}_i\to\vec{0}
~,
\end{equation}
%
or one parton to become soft
%
\begin{equation}
p_r\to\mathrm{soft} \quad\Leftrightarrow\quad E_r\to0
~.
\end{equation}
%
The jet function behaves in those limits such that
%
\begin{align}
\JetR\big(\pSetNpls\big)
&\overset{p_r\to\mathrm{soft}}{-\hspace{-0.2ex}-\hspace{-1.0ex}\longrightarrow}
\JetB\big(\pSetNOr\big)
\quad,
\\
\JetR\big(\pSetNpls\big)
&\overset{p_r\parallel p_i}{-\hspace{-0.2ex}-\hspace{-1.0ex}\longrightarrow}
\JetB\big(\pSetNOri\big)
\quad,
\\
\JetR\big(\pSetNpls\big)
&\overset{p_r\parallel\pP,\pM}{-\hspace{-0.2ex}-\hspace{-1.0ex}\longrightarrow}
\JetB\big(\pSetNOr\big)
\quad,
\end{align}
%
where 
%
%
\begin{equation}
\textit{
$\pSetNOr$ is obtained from $\pSetNpls$ by removing momentum $p_r$~,
}
\end{equation}
%
and
%
\begin{equation}
\textit{
$\pSetNOri$ is obtained by additionally replacing $p_i$ with $(1+z_{ri})p_i$
}
\end{equation}
%
with
%
\begin{equation}
z_{ri} = \frac{E_r}{E_i}
~.
\end{equation}
Of course, at the collinear limit $p_r\parallel p_i$, we have $(1+z_{ri})p_i=p_r+p_i$.
We silently assume here again that final-state parton momenta are light-like.

Let us consider the matrix element for collinear factorization, and imagine $\kperp=0$.
It is singular in these limits, but the singular behavior factorizes in a universal, and well-known, way.
We want to write them in a factorized form and level of detail necessary for the further discussion. 
Also, we would like to write them in a functional sense, such that the symbols representing momenta are identical on the left-hand side and the right-hand side. 
%
%
We have
%
\begin{align}
\Mtree{}\big(\Qin;\pSetNpls\big)
&\overset{p_r\to\mathrm{soft}}{-\hspace{-0.2ex}-\hspace{-1.0ex}\longrightarrow}
\hatRtimesA{\soft}{}{p_r}{\Qin;T^{\soft}_{r}\pSetNOr}
\quad,
\\
\Mtree{}\big(\Qin;\pSetNpls\big)
&\overset{p_r\parallel p_i}{-\hspace{-0.2ex}-\hspace{-1.0ex}\longrightarrow}
\hatRtimesA{\Fcol}{ir}{p_r}{\Qin;T^{\Fcol}_{r,i}\pSetNOri}
\quad,
\label{Eq:fincollim}
\\
\Mtree{}\big(\Qin;\pSetNpls\big)
&\overset{p_r\parallel\pP}{-\hspace{-0.2ex}-\hspace{-1.0ex}\longrightarrow}
\hatRtimesA{\Icol}{\inlbl r}{p_r}{\Qin-\xP_r\pP;T^{\Icol}_{r,\inlbl}\pSetNOr}
\\
\Mtree{}\big(\Qin;\pSetNpls\big)
&\overset{p_r\parallel\pM}{-\hspace{-0.2ex}-\hspace{-1.0ex}\longrightarrow}
\hatRtimesA{\Icol}{\inbar r}{p_r}{\Qin-\xM_r\pM;T^{\Icol}_{r,\inbar}\pSetNOr}
\quad.
\end{align}
%
The quantities $\hat{\EuScript{R}}$ represent the singular behavior as function of the radiative momentum $p_r$.
The quantities $\hat{\EuScript{A}}$ involve squares of tree-level scattering amplitudes with $n$ final-state momenta, but including color and spin correlations.
The mappings $T$ establish the aforementioned demand that the relations hold in the functional sense, and make sure that on-shellness and momentum conservation hold for the arguments of $\hat{\EuScript{A}}$, while the symbols represent the same momenta on the left-hand side and the right-hand side of the relations.
The sets of final-state momenta $\pSetNOr$ and $\pSetNOri$ do not satisfy momentum conservation, that is their sum is not equal momentum before the semicolon, and the mappings $T$ fix that by deforming them a bit, keeping them however on-mass shell \eg:
%
\begin{equation}
Q^\mu 
= \sum_{j=1}^{n+1}p_j^\mu
= \sum_{j=1}^n\big(T^{\soft}_{r}p_j\big)^\mu
\quad,\quad
\big(T^{\soft}_{r}p_j\big)^\mu\big(T^{\soft}_{r}p_j\big)_\mu = m_j^2
\quad.
\end{equation}
%
They could for example be the phase space mappings employed in the well-known subtraction schemes to make sure that the subtraction terms are valid at least in a non-vanishing phase space region around the singularities.
The limits for the matrix element are given in more detail by
%
\begin{align}
&\hatRtimesA{\soft}{}{p_r}{\Qin;T^{\soft}_{r}\pSetNOr}
\\&\hspace{20ex}=
-\frac{4\pi\alphaS}{\mu^\vep}
   \sum_{a,b\neq r}
   \frac{(\lop{n_a}{n_b})}{(\lop{n_a}{p_r})(\lop{p_r}{n_b})}
   \,\MtreeCor{a}{b}\big(\Qin;T^{\soft}_{r}\pSetNOr\big)
\notag~,\\
&\hatRtimesA{\Fcol}{ir}{p_r}{\Qin;T^{\Fcol}_{r,i}\pSetNOri}
\\&\hspace{20ex}=
\frac{4\pi\alphaS}{\mu^\vep}
  \,\frac{1}{\lop{p_i}{p_r}}
  \,\EuScript{Q}_{ir}(z_{ri})
    \otimes\Mtree{ir}\big(\Qin;T^{\Fcol}_{r,i}\pSetNOri\big)
\notag~,\\
&\hatRtimesA{\Icol}{\inlbl r}{p_r}{\Qin-\xP_r\pP;T^{\Icol}_{r,\inlbl}\pSetNOr}
\\&\hspace{20ex}=
  \frac{4\pi\alphaS}{\mu^\vep}
  \,\frac{-2}{\sTot\xP\xM_r}
  \,\EuScript{Q}_{\inlbl r}(-\xP_r/\xP)
    \otimes\Mtree{\inlbl r}\big(\Qin-\xP_r\pP;T^{\Icol}_{r,\inlbl}\pSetNOr\big)
\notag~.
\end{align}
%
Here, $\MtreeCor{a}{b}$ refers to the color-correlated squared matrix element for the process with one final-state gluon fewer than the original process, see \Appendix{App:correlators}.
The labels `$ir$' and `$\inlbl r$' for the other matrix elements indicate that it can be for a process with something less trivial than one final-state gluon fewer, like a quark-antiquark pair replaced by a gluon, an initial-state gluon replaced by quark and a final-state antiquark removed, etc.
The collinear splitting functions $\EuScript{Q}$ are defined in \Appendix{App:splittingfunctions} as functions of the ratio of the momentum fractions of the momenta becoming collinear.
It is understood that their subscripts refer to the flavors of the indicated particles in the process.
The minus-sign for the initial-state case stems from the fact that we define the splitting functions for matrix elements with all momenta outgoing, implying that $\lop{p_i}{p_r}$ is negative if $p_i$ is an initial-state momentum.
The symbol $\otimes$ for the collinear cases indicates that there are spin-correlated matrix elements involved, see \Appendix{App:correlators}.
For the initial-state collinear limit, we prefer to write the expression in terms of the variables $\xP_r,\xM_r$.
The $\inbar$-case has the same form with the bar-and non-bar quantities exchanged.
The $a$-sum and $b$-sum for the soft limit are over all external partons, both initial-state and final-state.

The singular behavior causes the real radiation integral, with $\Mtree{}\big(\Qin;\pSetNpls\big)$ and $\JetR\big(\pSetNpls\big)$, to be divergent, and the task at hand is to split it into a finite part and a divergent part that can be explicitly expressed as a Laurent expansion in $\vepv$ within dimensional regularization
%
\begin{equation}
\sigma_{\Real}(\vepv) = \sigma_{\Real}^{\mathrm{div}}(\vepv) + \sigma_{\Real}^{\mathrm{fin}} + \Ord(\vepv)
~.
\label{Eq:268}
\end{equation}
%
This decomposition is not unique, because $\sigma_{\Real}^{\mathrm{div}}(\vepv)$ will typically include finite pieces of $\Ord(\vepv^0)$.
Only the $1/\vepv^2$ and $1/\vepv$ terms in $\sigma_{\Real}^{\mathrm{div}}(\vepv)$ are universal.

The formulas for the singular limits in the foregoing are for the case when both initial-state momenta are light-like.
It turns out that if we assume the $\inlbl$-parton to be space-like, then the formulas for the limits are just the same, only with space-like matrix elements.
Despite non-vanishing $\kperp$, there is still a collinear singularity when a final-state gluon becomes collinear to the hadron momentum $\pP$ associated with the $\inlbl$-gluon.
It was shown in \cite{vanHameren:2022mtk} that the formulas for the singular limits have forms identical to the on-shell case, just with a particularly simple splitting function
%
\begin{equation}
\EuScript{Q}_{\inlbl r}(\zeta) = \frac{2C_g}{\zeta(1+\zeta)^2}
\quad\Leftrightarrow\quad
\EuScript{P}_{\inlbl r}(z)
\equiv
-z\EuScript{Q}_{\inlbl}(z-1) = \frac{2C_g}{z(1-z)}
~.
\end{equation}
%
In fact, there are no spin correlations involved and the symbol $\otimes$ is just multiplication in this case.
With this work we numerically confirm the correctness of this formula by observing that our subtraction terms indeed cure the singularity.

The sum of the $n$ final-state momenta must be equal to the total initial-state momentum, the quantities before the semicolon, in order for the matrix elements to be well-defined.
The mappings $T$ before involved only final-state momenta, as is the case in most subtraction schemes, but this is not apriori necessary, and the initial-state momenta could be involved in the re-distribution of the recoil too.
For collinear factorization, the initial-state momenta can only accomodate longitudinal momentum components.
For hybrid factorization, they can accomodate a full $4$-momentum, and we can simply re-distribute the recoil over the initial-state momenta as
%
%
\begin{align}
\Mtree{}\big(\Qin;\pSetNpls\big)
&\overset{p_r\to\mathrm{soft}}{-\hspace{-0.2ex}-\hspace{-1.0ex}\longrightarrow}
\hatRtimesA{\soft}{}{p_r}{\Qin-p_r;\pSetNOr}
\quad,
\\
\Mtree{}\big(\Qin;\pSetNpls\big)
&\overset{p_r\parallel p_i}{-\hspace{-0.2ex}-\hspace{-1.0ex}\longrightarrow}
\hatRtimesA{\Fcol}{ir}{p_r}{\Qin-p_r+z_{ri}p_i;\pSetNOri}
\quad,
\\
\Mtree{}\big(\Qin;\pSetNpls\big)
&\overset{p_r\parallel\pP/\pM}{-\hspace{-0.2ex}-\hspace{-1.0ex}\longrightarrow}
\hatRtimesA{\Icol}{\inlbl/\inbar,r}{p_r}{\Qin-p_r;\pSetNOr}
\;\;.
\end{align}
%
Because the expressions on the right-hand-side are well-defined all over phase space, and not only at the limit, they can be used to construct subtraction terms in order establish \Equation{Eq:268}.
The formulas above cannot be the {\em only} ingredient for the subtraction terms, because of double subtraction of soft-collinear singularities.
We also want to augment them with phase space restrictions limiting their support around the singularities they are supposed to subtract.
Finally, it will prove to be beneficial for soft terms with final-state collinear singularities to apply the final-state collinear type of phase space mapping.
Fact is, however, that subtraction terms can be cast into the two argument types $\big(\Qin-p_r+z_{ri}p_i;\pSetNOri\big)$ and $\big(\Qin-p_r;\pSetNOr\big)$.
We introduce new objects $\EuScript{R}$ and $\EuScript{A}$, now without hat, that will be precisely defined later, but fit the aforementioned forms.

\subsection{Organization of the subtraction terms}
We define the finite radiative integral as
%
\begin{align}
&\sigma_{\Real}^{\mathrm{fin}}=
\frac{1}{\symfac_{n+1}}\int\dQin
\int d\Phi\big(\Qin;\pSetNpls\big)
\bigg\{
\EL{\big(\Qin;\pSetNpls)\big)}
\,\Mtree{}\big(\Qin;\pSetNpls\big)
\,\JetR\big(\pSetNpls\big)
\label{Eq:229}\\&\hspace{54ex}
-
\sum_{r}\mathrm{Subt}_r\big(\Qin;\pSetNpls\big)
\bigg\}
\notag~,
\end{align}
where the $r$-sum is over all final-state partons, and with
%
\begin{align}
\mathrm{Subt}_r\big(\Qin;\pSetNpls\big)
&=
  \hspace{0.5ex}\sum_{i}
   \EL{\big(\Qin-q_{r,i};\pSetNOri\big)}
   \,\RtimesAarg{\mathrm{F}}{ir}{p_r}{\Qin-p_r+z_{ri}p_i;\pSetNOri}
   \,\JetB\big(\pSetNOri\big)
\notag\\&\hspace{0ex}
  +\hspace{-1ex}\sum_{a\in\{\inlbl,\inbar\}}\hspace{-1ex}\EL{\big(\Qin-q_{r,a};\pSetNOr\big)}
   \,\RtimesAarg{\Icol}{ar}{p_r}{\Qin-p_r;\pSetNOr}
   \,\JetB\big(\pSetNOr\big)
\label{Eq:345}\\&\hspace{0ex}
   +\hspace{-1ex}\sum_{a\in\{\inlbl,\inbar\}}\hspace{-1ex}\EL{\big(\Qin-q_r;\pSetNOr\big)}
   \,\RtimesAarg{\Isoft}{a}{p_r}{\Qin-p_r;\pSetNOr}
   \,\JetB\big(\pSetNOr\big)
\notag\\&\hspace{0ex}
   +\hspace{-1ex}\sum_{a\in\{\inlbl,\inbar\}}\hspace{-1ex}\EL{\big(\Qin-q_r;\pSetNOr\big)}
   \,\RtimesAarg{\Isoftcol}{a}{p_r}{\Qin-p_r;\pSetNOr}
   \,\JetB\big(\pSetNOr\big)
\notag~,
\end{align}
where also the $i$-sum is over all final-state partons with $\EuScript{R}^{\mathrm{F}}_{rr}(p_r)\equiv0$.
We make sure to avoid double counting in the precise definition of the terms.
The terms with label `F' refer to final-state singularities and the ones with `I' to initial-state singularities.
Some of the terms carry the label $r$ because they depend on the flavor of final-state $r$.
The first three lines categorize the three different types of momentum recoils that appear in our scheme.
The fourth and third are identical with respect to the recoil, but we decided to categorize them separately already for later convenience.

We subtract a momentum $q$, with subscripts depending on the subtraction term, from the arguments of the $\EL$-function.
This is allowed as long as this momentum vanishes in the limit for which the respective subtraction term is supposed to cure a singularity.
In our situation of $k_T$-factorization, this momentum can have both non-vanishing longitudinal {\em and} transverse components.
For the initial-state soft case, this could for example be $p_r$ itself.
For the initial-state collinear cases, this momentum {\em cannot} be equal to $p_r$, since it does not vanish in those collinear limits.
Then, it can be ``at most'' $q_{r,\inlbl}=\xM_r\pM+\kperp$ and $q_{r,\inbar}=\xP_r\pP+\kperp$.
For the other cases, the $q$-momentum can be identical to the recoil subtracted from the initial-state momenta in the matrix element.

Using the result of \Appendix{App:CollMap}, we find that the subtraction terms integrate to
\begin{align}
\sigma_{\Real}^{\mathrm{div}}(\vepv)
&=
\frac{1}{\symfac_{n+1}}\sum_r
\int\dQin\int d\Phi\big(\Qin;\pSetNOr\big)
\,\JetB\big(\pSetNOr\big)
\label{Eq:394}\\&\hspace{4ex}\times
\bigg\{
   \sum_i\LtimesAarg{\mathrm{F}}{ir}{\vepv,\Qin,\pSetNOr}{\Qin;\pSetNOr}
   +\hspace{-1ex}\sum_{a\in\{\inlbl,\inbar\}}
     \hspace{-1ex}\LtimesAarg{I}{ar}{\vepv,\Qin,\pSetNOr}{\Qin;\pSetNOr}
\bigg\}
~,
\notag
\end{align}
%
where
%
\begin{equation}
\LtimesA{I}{ar} =
 \LtimesA{\Icol}{ar}
+\LtimesA{\Isoft}{a}
+\LtimesA{\Isoftcol}{a}
~,
\end{equation}
%
with
%
\begin{align}
\EuScript{L}^{\mathrm{F}}_{ir}\big(\vepv,\Qin,\pSetNOr\big)
&= 
   \int\frac{d^{4-2\vepv}p_r}{(2\pi)^{3-2\vepv}}\,\delta_+(p_r^2)
  \,(1-z_{ri})\,\EuScript{R}^{\mathrm{F}}_{ir}(p_r)
\,\Theta(p_r-z_{ri}p_i)
\label{Eq:461}\\&\hspace{28ex}\times
\EL{\big(\Qin+p_r-z_{ri}p_i-q_{r,i};\pSetNOr\big)}
~,\notag\\
\EuScript{L}^{\Icol}_{ar}\big(\vepv,\Qin,\pSetNOr\big)
&= \int\frac{d^{4-2\vepv}p_r}{(2\pi)^{3-2\vepv}}\,\delta_+(p_r^2)
  \,\EuScript{R}^{\Icol}_{ar}(p_r)
\,\Theta(p_r)
  \,\EL{\big(\Qin+p_r-q_{r,a};\pSetNOr\big)}
\label{Eq:411}
~,\\
\EuScript{L}^{\Isoftsoco}_{a}\big(\vepv,\Qin,\pSetNOr\big)
&= 
   \int\frac{d^{4-2\vepv}p_r}{(2\pi)^{3-2\vepv}}\,\delta_+(p_r^2)
  \,\EuScript{R}^{\Isoftsoco}_{a}(p_r)
\,\Theta(p_r)
\,\EL{\big(\Qin+p_r-q_r;\pSetNOr\big)}
~.\label{Eq:542}
\end{align}
%
and
%
\begin{equation}
\Theta(q) = \theta(-\xP<\xP_q<1-\xP)\,\theta(-\xM<\xM_q<1-\xM)
\label{Eq:574}~.
\end{equation}
%
From these formulas it appears that the best choice for the $q$-momenta is it to be equal to the recoil, so the $\EL$-function does not depend on the integration variables of $p_r$ anymore.
As mentioned before, this is not possible for the initial-state collinear cases, and we cannot avoid arguments $\EL{\big(\Qin+\xP_r\pP;\pSetNOr\big)}$ and $\EL{\big(\Qin+\xM_r\pM;\pSetNOr\big)}$.
This of course also happens in the well-known subtraction formulas for collinear factorization, leading to what in~\cite{Catani:1996vz} is called the ``P-operator''.

While the $\vepv^{-2}$- and cancelling $\vepv^{-1}$-parts {\em can} be obtained analytically by applying a subtraction approach on these integrals again, the finite remainder of the integrals cannot be calculated completely analytically.
This is relevant also when we choose the $q$-momentum to be equal to the recoil, seemingly removing the $\EL$-function from the integral, because the extraction of the $\vepv$ poles will typically not follow the integration borders dictated by $\Theta(q)$ in order to be performed analytically.

Another issue is that, while formally correct, subtracting the recoil from $Q$ inside the $\EL$ function may not work very well in practice.
The reason is that, although the recoil may be small, it is not guaranteed that its transverse components are small compared to the initial-state $\kperp$.
The longitudinal variables $\xP,\xM$ are upper limits on the size of the longitudinal components of the recoil, so the latter can indeed be considered small, but for the transverse components this is not the case.
%
%

Whether this becomes a problem depends on the actual behavior of the $\kperp$-dependent PDF.
If it does, then the solution is to not subtract the transverse components of the recoil from $\Qin$ inside the $\EL{}$ function, with the consequence for all integrated subtraction terms the finite part requires numerical integration with the $\EL{}$ function as part of the integrand.
Realize, however, that these numerical integrals can be performed in a Monte Carlo approach, by including them into the phase space integral.
In our actual numerical studies, we used the parton-branching $k_T$-dependent PDFs~\cite{BermudezMartinez:2018fsv,Abdulov:2021ivr}, and the aforementioned problem did not occur.
Still, in the following, we will work out the integrals for the safe, ``conservative'', approach.
We will express them in terms of divergent integrals that are calculated analytically, and are labelled `$\mathrm{div}$', and finite integrals labelled `$\mathrm{fin}$' that must be integrated numerically.
In the ``progressive'' approach, the non-trivial integrands reduce to theta-functions only enforcing integration limits. 
More precisely, the difference between the two approaches is expressed by the function
%
\begin{equation}
\ELshiftPerp{q}
=
\left\{
\begin{array}{ll}
  \displaystyle\frac{\EL{\big(\Qin+q_{\theperp};\pSetN\big)}}{\EL{\big(\Qin;\pSetN\big)}}
  =\frac{F_{\inlbl}\big(\xP,\kperp+q_{\theperp},\muF(\pSetN)\big)}
     {F_{\inlbl}\big(\xP,\kperp,\muF(\pSetN)\big)}
       & \textrm{conservative approach} \\
      1 & \textrm{progressive approach}\\
\end{array} 
\right. 
\label{Eq:1043}
\end{equation}
%
that will appear in the integrals of the subtraction terms.

\subsection{Definition of the subtraction terms}
Now we present the actual subtraction terms more precisely.
These follow the same structure as in~\cite{Somogyi:2006cz,Somogyi:2009ri}, but with different phase space mappings, and organized into different groups.
It turns out to be beneficial to apply the well-known technique of splitting the soft factor into two terms each with only one collinear singularity
\begin{equation}
\frac{(\lop{n_a}{n_b})}{(\lop{n_a}{p_r})(\lop{p_r}{n_b})}
=
\frac{1}{\lop{n_a}{p_r}}\,
\frac{\lop{n_a}{n_b}}{\lop{n_a}{p_r}+\lop{p_r}{n_b}}
+
\frac{1}{\lop{p_r}{n_b}}\,
\frac{\lop{n_a}{n_b}}{\lop{n_a}{p_r}+\lop{p_r}{n_b}}
~.
\end{equation}
%
For each of these with final state $a=i$, we can then apply the final-state collinear type of mapping, adding the energy of the radiation to the radiator.
We also mention that for the final-state collinear terms, we avoid double counting with the help of a selector function $\theta(E_r<E_i)$ demanding that a radiator has larger energy than the radiation.

\subsubsection{Final-state terms}
Firstly, there are the terms in \Equation{Eq:345} with arguments $\big(\Qin-p_r+z_{ri}p_i;\pSetNOri\big)$.
While they do involve initial-state ``spectators'' in the soft terms, we refer to them collectively as ``final-state terms''.
They are given by 
%
\begin{equation}
\RtimesA{\mathrm{F}}{ir}
=
\RtimesA{\Fcol}{ir}
+\RtimesA{\Fsoft}{i}
+\RtimesA{\Fsoftcol}{i}
\label{Eq:540}~,
\end{equation}
%
with
%
\begin{align}
\RtimesA{\Fcol}{ir}
&=
\frac{4\pi\alphaS}{\mu^\vep}
  \,\theta(\lop{n_r}{n_i}<2\zeta_0)
  \,\frac{\theta(E_r<E_i)}{\lop{p_i}{p_r}}
  \,\EuScript{Q}_{ir}(z_{ri})
    \otimes\Mtree{ir}
\label{Eq:612}~,
\\
\RtimesA{\Fsoft}{i}
&=
-\frac{4\pi\alphaS}{\mu^\vep}
  \,\theta(E_r<E_0)
  \,\frac{2}{\lop{n_i}{p_r}}
  \sum_b
  \frac{\lop{n_i}{n_b}}{\lop{n_i}{p_r}+\lop{n_b}{p_r}}
  \,\MtreeCor{i}{b}
\label{Eq:613}~,
\\
\RtimesA{\Fsoftcol}{i}
&=
-\frac{4\pi\alphaS}{\mu^\vep}
 \,\theta(E_r<E_0)\theta(\lop{n_r}{n_i}<2\zeta_0)
  \,\frac{2C_i}{\lop{p_i}{p_r}}\,\frac{1}{z_{ri}}
  \,\Mtree{}
\end{align}
%
All these matrix elements carry the aforementioned arguments. 
The $b$-sum is over all partons, initial-state and final-state (and $b\neq i$, but then again $\lop{n_i}{n_i}=0$).
The third line contains the soft-collinear counter term correcting the double subtraction of soft-collinear singularities.
We use the notation
%
\begin{equation}
C_i=
\left\{
\begin{array}{ll}
  C_g=C_A=\Nc &\quad\textrm{if $i$ refers to a gluon,}\\
  C_q=C_F=(\Nc-1/\Nc)/2 &\quad\textrm{if $i$ refers to a quark or antiquark,}\\
\end{array} 
\right.
\end{equation}
%
and later we will also use
%
\begin{equation}
C_{ij}=
\left\{
\begin{array}{ll}
  C_g &\quad\textrm{if both $i,j$ refer to gluons,}\\
  C_q &\quad\textrm{if exactly one of $i,j$ refers to a gluon,}\\
  T_R &\quad\textrm{if none of $i,j$ refers to a gluon.}\\
\end{array} 
\right.
\end{equation}
%
Color conservation implies
%
\begin{equation}
\sum_{b\neq i}\MtreeCor{i}{b} = -C_i\,\Mtree{}~.
\end{equation}
%
The soft and soft-collinear terms vanish if $r$ does not refer to a gluon.
The restriction $\theta(E_r<E_0)$ appears both for the soft and the soft-collinear term, so these clearly match in the collinear limit $n_r\to n_i$, undoing the soft-collinear double subtraction.
In the soft limit $E_r\to0$, the restriction $\theta(E_r<E_i)=1$, so then the collinear and the soft-collinear term match.
The final result for $\sigma_{\Real}$ should not depend on the the parameters $E_0,\zeta_0$ which in practice is a powerful check of correctness.
Furthermore, it opens the possibility to adjust the behavior of the integrand under numerical integration, and we find that it is improved by the restrictions.
%
Also, for smaller values of the restricting parameters, fewer subtraction terms contribute to a given phase space point, making the evaluation per phase space computationally point cheaper.
In particular, by choosing $E_0=E_i$ (or $\min(E_0,E_i))$, at most $1$ of the $2$ assignments ``radiator'' versus ``radiation'' for $i,r$ in the soft terms contributes per phase space point.

\subsubsection{Initial-state terms}
The terms in \Equation{Eq:345} with arguments $\big(\Qin-p_r;\pSetNOr\big)$ have initial-state radiators $(\inlbl,\inbar)$ instead of the final-state $i$ in the previous terms.
We only present the $\inlbl$-terms.
The $\inbar$-terms are completely analogous with exchange of the bar-and non-bar variables.
Due to universality they largely fit the same formulas as the final-state ones, but we find it more convenient to write the collinear terms using the variables $\xM_r=2\lop{\pP}{p_r}/\sTot$ and $\xP_r=2\lop{\pM}{p_r}/\sTot$:
%
\begin{align}
\RtimesA{\Icol}{\inlbl r} &= 
  \frac{4\pi\alphaS}{\mu^\vep}
  \theta\big(\xM_r<\xiP\xP_r\big)
  \,\frac{-2}{\sTot\xM_r\xP}
  \,\EuScript{Q}_{\inlbl r}(-\xP_r/\xP)
    \otimes\Mtree{\inlbl r}
\label{Eq:1040}
\\
\RtimesA{\Isoft}{\inlbl} &= 
  -\frac{4\pi\alphaS}{\mu^\vep}
  \,\theta(E_r<E_0)
  \,\frac{2}{\lop{n_{\inlbl}}{p_r}}
  \sum_b
  \frac{\lop{n_{\inlbl}}{n_b}}{\lop{n_{\inlbl}}{p_r}+\lop{n_b}{p_r}}
  \,\MtreeCor{\inlbl}{b}
\label{Eq:983}
\\
\RtimesA{\Isoftcol}{\inlbl} &= 
  -\frac{4\pi\alphaS}{\mu^\vep}
  \,\theta(E_r<E_0)\theta\big(\xM_r<\xiP\xP_r\big)
  \,\frac{4C_{\inlbl}}{\sTot\xP_r\xM_r}
  \,\Mtree{}
~.\label{Eq:728}
\end{align}
%
There is only a collinear singularity for the space-like $\inlbl$ case if the radiation is a gluon, so no subtraction term with a quark or antiquark as radiation is required.
For the on-shell $\inbar$-terms, they are included.

\section{Calculation of the integrated subtraction terms}
In the following, we present the quantities $\LXYZ{X,Y}{Z}$ in the decomposition of the integrated subtraction terms of \Equation{Eq:394} as
%
\begin{align}
\LtimesA{\mathrm{F}}{ir}
&=
 \cNLO\,\EL{\big(\Qin;\pSetN\big)}\bigg[
  \LXYZ{\Fcol}{ir}\Mtree{ir}
   +\sum_b\LXYZ{\Fsoft}{ib}\MtreeCor{i}{b}
  +\LXYZ{\Fsoftcol}{i}\Mtree{}
 \bigg]
~,
\\
\LtimesA{I}{ar} &= 
 \cNLO\,\EL{\big(\Qin;\pSetN\big)}\bigg[
  \LXYZ{\Icol}{ar}\Mtree{ar}
  +\sum_b\LXYZ{\Isoft}{a,b}\MtreeCor{a}{b}
  +\LXYZ{\Isoftcol}{a}\Mtree{}
 \bigg]
~.
\end{align}
%
Each of the $\LXYZ{X,Y}{Z}(\vepv)$ will then be decomposed into a divergent part that can be expanded into a Laurent series in $\vepv$ analytically, and a finite part that must be integrated numerically
\begin{equation}
\LXYZ{X,Y}{Z}(\vepv)
=
\LXYZ{X,Y,\mathrm{div}}{Z}(\vepv)
+\LXYZ{X,Y,\mathrm{fin}}{Z}
+\Ord(\vepv)
~.
\end{equation}
%
Just to be clear, we mention that $\EL{\big(\Qin;\pSetN\big)}$ contains the PDFs for the original radiative process, which may indeed be the ``wrong'' ones for processes of $\Mtree{ar}$, in which say an initial-state gluon may have been replaced by a quark.
This is usually the case for any subtraction scheme, but not always so explicitly visible if the formulas are written only at the parton-level.

\subsection{Final-state terms}
We need to combine \Equation{Eq:461} and \Equation{Eq:540}.
We remind the reader of the functions $\Theta(q)$ in \Equation{Eq:574} and $\ELshiftPerp{q}$ in \Equation{Eq:1043}, which are defined applying the decomposition of the momentum $q$ in \Equation{Eq:18}.

\subsubsection{Collinear terms}
Realize that the explicit $p_i$ and $z_{ri}$ in \Equation{Eq:612} are the ones before changeing integration variables from $p_i$ to $\tilde{p}_i=(1+z_{ri})p_i$, so for the integrated subtraction terms we need to insert $z_{ri}=\tilde{z}_{ri}/(1-\tilde{z}_{ri})$ and $p_i=(1-\tilde{z}_{ri})\tilde{p}_i$, and then just $\tilde{p}_i\overset{\mathrm{rename}}{\longrightarrow}p_i$, $\tilde{z}_{ri}\overset{\mathrm{rename}}{\longrightarrow}z_{ri}$.
%
\begin{align}
\LXYZ{\Fcol}{ir}
&=
\frac{1}{\piep\mu^\vep}\int d^{4+\vep}p_r\delta_+(p_r^2)(1-z_{ri})
\,\ELshiftPerp{p_r-z_{ri}p_i}\,\Theta(p_r-z_{ri}p_i)
\notag\\&\hspace{12ex}\times
\theta(\lop{n_r}{n_i}<2\zeta_0)
\,\theta\bigg(\frac{1-2z_{ri}}{1-z_{ri}}\bigg)
\,\frac{1}{(1-z_{ri})\lop{p_i}{p_r}}
\,\EuScript{P}_{ir}(1-z_{ri})
&
\notag\\&\hspace{0ex}=
\frac{1}{\piep\mu^\vep}\int d^{4+\vep}p_r\delta_+(p_r^2)
\,\SXYZ{\Fcol}{}(p_r;p_i)
\,\ELshiftPerp{p_r-z_{ri}p_i}\,\Theta(p_r-z_{ri}p_i)
\end{align}
%
with
%
\begin{equation}
\SXYZ{\Fcol}{}(p_r;p_i)
=
\frac{\theta(\lop{n_r}{n_i}<2\zeta_0)\,\theta(z_{ri}<\srac{1}{2})}
       {\lop{p_i}{p_r}}
\,\EuScript{P}_{ir}(1-z_{ri})
~.
\end{equation}
%
The function $\SXYZ{\Fcol}{}$ is singular both in the collinear limit $\vec{n}_r\to\vec{n}_i$, and in the soft limit $E_r\to0$ because of the $1/z_{ri}$ hidden in $\EuScript{P}_{ir}(1-z_{ri})$.
In either limit, however, $p_{r\theperp}-z_{ri}p_{i\theperp}$ vanishes.
Thus we see that the integral
%
\begin{align}
\LXYZ{\Fcolfin}{ir}
&=
\frac{1}{\pi}
\int d^{4}p_r\,\delta_+(p_r^2)
\,\SXYZ{\Fcol}{}(p_r;p_i)
\,\Big[\ELshiftPerp{p_r-z_{ri}p_i}\,\Theta(p_r-z_{ri}p_i)-1\Big]
\end{align}
%
is finite, and
%
\begin{align}
\LXYZ{\Fcoldiv}{ir}(\vepv)
&=
\frac{1}{\piep\mu^{\vep}}
\int d^{4+\vep}p_r\,\delta_+(p_r^2)
\,\SXYZ{\Fcol}{}(p_r;p_i)
\end{align}
%
can be calculated analytically.

\subsubsection{Soft terms}
Again because $p_{r\theperp}-z_{ri}p_{i\theperp}$ vanishes both in the soft and the collinear limit, the integral
%
\begin{align}
\LXYZ{\Fsoftfin}{ib,\mathrm{compl}}
&=
\frac{-2}{\pi}
\int d^{4}p_r\,\delta_+(p_r^2)
\,\frac{1}{\lop{n_i}{p_r}}\,\frac{\lop{n_i}{n_b}}{\lop{n_i}{p_r}+\lop{n_b}{p_r}}
\,\theta(E_r<E_0)\,(1-z_{ri})
\notag\\&\hspace{36ex}\times
\Big[\ELshiftPerp{p_r-z_{ri}p_i}\,\Theta(p_r-z_{ri}p_i)-1\Big]
\end{align}
%
is finite, while
%
\begin{align}
\LXYZ{\Fsoftdiv}{ib,\mathrm{compl}}(\vepv)
&=
\frac{-2}{\piep\mu^{\vep}}
\int d^{4+\vep}p_r\,\delta_+(p_r^2)
\,\frac{1}{\lop{n_i}{p_r}}\,\frac{\lop{n_i}{n_b}}{\lop{n_i}{p_r}+\lop{n_b}{p_r}}
\,\theta(E_r<E_0)\,(1-z_{ri})
\end{align}
%
can, in principle, be calculated analytically.
We find it however too complicated.
Instead, we introduce
%
\begin{equation}
E_r^{(ib)}
=
\frac{\lop{n_b}{p_r}}{\lop{n_i}{n_b}}
+
\frac{\lop{n_i}{p_r}}{\lop{n_i}{n_b}}
=
E_r\,\frac{\lop{n_r}{n_b}+\lop{n_i}{n_r}}{\lop{n_i}{n_b}}
\end{equation}
%
which vanishes in the soft limit, and becomes equal to $E_r$ in the collinear limit, so we can define
%
\begin{align}
&\LXYZ{\Fsoftfin}{ib}
=
\frac{-2}{\pi}
\int d^{4}p_r\,\delta_+(p_r^2)
\,\frac{1}{\lop{n_i}{p_r}}\,\frac{\lop{n_i}{n_b}}{\lop{n_i}{p_r}+\lop{n_b}{p_r}}
\notag\\&\hspace{0ex}\times
\bigg[\ELshiftPerp{p_r-z_{ri}p_i}\,\Theta(p_r-z_{ri}p_i)
     \,\theta\big(E_r<E_0\big)\,\bigg(1-\frac{E_r}{E_i}\bigg)
     -\theta\big(E_r^{(ib)}<E_0\big)\,\bigg(1-\frac{E_r^{(ib)}}{E_i}\bigg)
\bigg]
\end{align}
%
with
%
\begin{align}
\LXYZ{\Fsoftdiv}{ib}(\vepv)
&=
\frac{-2}{\piep\mu^{\vep}}
\int d^{4+\vep}p_r\,\delta_+(p_r^2)
\,\frac{1}{\lop{n_i}{p_r}}\,\frac{\lop{n_i}{n_b}}{\lop{n_i}{p_r}+\lop{n_b}{p_r}}
\,\theta\big(E_r^{(ib)}<E_0\big)\,\bigg(1-\frac{E_r^{(ib)}}{E_i}\bigg)
~,
\end{align}
%
which is easier to calculate analytically.

\subsubsection{Soft-collinear terms}
Analogously, we have
\begin{align}
\LXYZ{\Fsoftcolfin}{i}
&=
\frac{-2C_i}{\pi}
\int d^{4}p_r\,\delta_+(p_r^2)
\,\SXYZ{\Fsoftcol}{}(p_r;p_i)
\,\Big[\ELshiftPerp{p_r-z_{ri}p_i}\,\Theta(p_r-z_{ri}p_i)-1\Big]
\\
\LXYZ{\Fsoftcoldiv}{i}(\vepv)
&=
\frac{-2C_i}{\piep\mu^{\vep}}
\int d^{4+\vep}p_r\,\delta_+(p_r^2)
\,\SXYZ{\Fsoftcol}{}(p_r;p_i)
\end{align}
%
with
%
\begin{align}
\SXYZ{\Fsoftcol}{}(p_r;p_i)
&=
\theta(E_r<E_0)\theta(\lop{n_i}{n_r}<2\zeta_0)(1-z_{ri})
\,\frac{1}{\lop{p_i}{p_r}}\,\frac{1}{z_{ri}}
~.
\end{align}
\subsection{Initial-state soft and soft-collinear terms}
We need to combine \Equation{Eq:542} with \Equation{Eq:983} and \Equation{Eq:728}.
We will only handle the $\inlbl$-terms.
The $\inbar$-terms are completely analogous with exchange of the bar-and non-bar variables.
The arguments $\xP_q=\xP_r,\xM_q=\xM_r$ of $\Theta(q)$ in \Equation{Eq:574} now stay positive.
Furthermore, we will pick this function apart into its two factors and write those explicitly instead of the symbol $\Theta$.

\subsubsection{Soft}
Let us introduce the abbreviation
%
\begin{equation}
[dV(p_r,\vepv)] 
=
\frac{-2}{\piep\mu^\vep}\,d^{4+\vep}p_r\,\delta_+(p_r^2)
\,\frac{1}{E_r^2\,\lop{n_{\inlbl}}{n_r}}
~.
\end{equation}
%
We need to calculate
%
\begin{equation}
\LXYZ{\Isoft}{\inlbl b}(\vepv)
=
\int[dV(p_r,\vepv)]
\,\frac{\lop{n_{\inlbl}}{n_b}}{\lop{n_r}{n_{\inlbl}}+\lop{n_r}{n_b}}
\,\ELshiftPerp{p_{r}}\,\theta(\xP_r<1-\xP)\theta(\xM_r<1-\xM)\,\theta(E_r<E_0)
~,
\end{equation}
with $\ELshiftPerp{q}$ as in \Equation{Eq:1043}.
Since $p_{r\theperp}$ vanishes both in the soft and the initial-state collinear limit, we see again that a single subtraction suffices and that we can define
%
\begin{align}
\LXYZ{\Isoftfin,1}{\inlbl b}
&=
\int[dV(p_r,0)]
\,\frac{\lop{n_{\inlbl}}{n_b}}{\lop{n_r}{n_{\inlbl}}+\lop{n_r}{n_b}}
\,\theta(\xP_r<1-\xP)
\notag\\&\hspace{16ex}\times
\Big[
  \ELshiftPerp{p_{r}}\theta(\xM_r<1-\xM)\,\theta(E_r<E_0) - \theta\big(E_r^{(\inlbl b)}<E_0\big)
\Big]
\\
\LXYZ{\Isoftdiv,1}{\inlbl b}(\vepv)
&=
\int[dV(p_r,\vepv)]
\,\frac{\lop{n_{\inlbl}}{n_b}}{\lop{n_{\inlbl}}{n_r}+\lop{n_b}{n_r}}
\,\theta(\xP_r<1-\xP)
\,\theta\big(E_r^{(\inlbl b)}<E_0\big)
~.
\end{align}
%
We gave the quantities an extra label since they are not the final ones.
We find the integral for $\LXYZ{\Isoftdiv,1}{\inlbl b}(\vepv)$ too complicated because it both depends on $n_b$ and includes the restriction $\theta(\xP_r<1-\xP)$.
We split it further into
%
\begin{equation}
\LXYZ{\Isoftdiv,1}{\inlbl b}(\vepv)
=
\LXYZ{\Isoftdiv}{\inlbl b}(\vepv) 
+ \LXYZ{\Isoftdiv,2}{\inlbl b}(\vepv)
+ \LXYZ{\Isoftfin,2}{\inlbl b} + \Ord(\vepv)
\end{equation}
%
with
%
\begin{align}
&\LXYZ{\Isoftfin,2}{\inlbl b}
=
\int[dV(p_r,0)]
\bigg[
  \theta(1-\xP<E_r/E)\,\theta(E_r<E_0)
\notag\\&\hspace{24ex}
 -\frac{\lop{n_{\inlbl}}{n_b}}{\lop{n_{\inlbl}}{n_r}+\lop{n_b}{n_r}}
    \,\theta(1-\xP<\xP_r)\,\theta\big(E_r^{(\inlbl b)}<E_0\big)
\bigg]
\\
&\LXYZ{\Isoftdiv}{\inlbl b}(\vepv)
=
\int[dV(p_r,\vepv)]
\,\frac{\lop{n_{\inlbl}}{n_b}}{\lop{n_{\inlbl}}{n_r}+\lop{n_b}{n_r}}
\,\theta\big(E_r^{(\inlbl b)}<E_0\big)
\\
&\LXYZ{\Isoftdiv,2}{\inlbl b}(\vepv)
=
-
\int[dV(p_r,\vepv)]\,\theta(1-\xP<E_r/E)\,\theta(E_r<E_0)
~.
\end{align}
%
The $\theta$ functions in $\LXYZ{\Isoftfin,2}{\inlbl b}$ prevent $p_r$ from becoming soft, while they become identical in the collinear limit $n_r\to n_{\inlbl}\Leftrightarrow\xM_r\to0$.
Notice that $\LXYZ{\Isoftdiv,2}{\inlbl b}(\vepv)$ does not depend on the spectator $b$, and will prove to combine nicely with a similar soft-collinear contribution.
The final finite soft contribution we denote
%
\begin{equation}
\LXYZ{\Isoftfin}{\inlbl b} = \LXYZ{\Isoftfin,1}{\inlbl b} + \LXYZ{\Isoftfin,2}{\inlbl b}
~.
\end{equation}

\subsubsection{Soft-collinear}
For the soft-collinear case, we abbreviate
\begin{equation}
[dW(p_r,\vepv)] = \,\frac{-2C_{\inlbl}}{\piep\mu^\vep}\,\frac{2}{\sTot}
\,d^{4+\vep}p_r\,\delta_+(p_r^2)
\,\frac{\theta(E_r<E_0)\,\theta(\xM_r<\xiP\xP_r)}{\xP_r\xM_r}
~.
\end{equation}
%
We need to calculate
%
\begin{align}
\LXYZ{\Isoftcol}{\inlbl}(\vepv)
&=
\int[dW(p_r,\vepv)]
\,\ELshiftPerp{p_{r}}
\,\theta(\xP_r<1-\xP)\theta(\xM_r<1-\xM)
~,
\end{align}
%
and split it as
\begin{align}
\LXYZ{\Isoftcolfin,1}{\inlbl}
&=
\int[dW(p_r,0)]\,\theta(\xP_r<1-\xP)
\Big[\ELshiftPerp{p_{r}}\,\theta(\xM_r<1-\xM) -1\Big]
\\
\LXYZ{\Isoftcoldiv,1}{\inlbl}(\vepv)
&=
\int[dW(p_r,\vepv)]\,\theta(\xP_r<1-\xP)
\\
&=
\LXYZ{\Isoftcoldiv}{\inlbl}(\vepv) + \LXYZ{\Isoftcoldiv,2}{\inlbl}(\vepv) + \LXYZ{\Isoftcolfin,2}{\inlbl} + \Ord(\vepv)
~,
\end{align}
%
with
%
\begin{align}
\LXYZ{\Isoftcolfin,2}{\inlbl}
&=
\int[dW(p_r,0)]\Big[\theta(1-\xP<E_r/E)-\theta(1-\xP<\xP_r)\Big]
\\
\LXYZ{\Isoftcoldiv}{\inlbl}(\vepv)
&=
\int[dW(p_r,\vepv)]
\\
\LXYZ{\Isoftcoldiv,2}{\inlbl}(\vepv)
&=
-\int[dW(p_r,\vepv)]\,\theta(1-\xP<E_r/E)
~.
\end{align}
%
By explicit calculation we see that, up to the color factor, the $1/\vepv$ pole in $\LXYZ{\Isoftcoldiv,2}{\inlbl}(\vepv)$ is identical to the one in $\LXYZ{\Isoftdiv,2}{\inlbl}(\vepv)$, so with the help of color conservation it will cancel in the sum over spectators for the latter.
We collect the remaining finite pieces in
%
\begin{equation}
\LXYZ{\Isoftcolfin}{\inlbl}
=
\LXYZ{\Isoftcolfin,1}{\inlbl}
+
\LXYZ{\Isoftcolfin,2}{\inlbl}
+
\Big[\LXYZ{\Isoftcoldiv,2}{\inlbl}(\vepv)-C_{\inlbl}\LXYZ{\Isoftdiv,2}{\inlbl}(\vepv)\Big]_{\vepv\to0}
~.
\end{equation}

\subsection{Initial-state collinear terms}
We need to combine \Equation{Eq:411} and \Equation{Eq:1040}.
%
Realize that the variables $\xP,\xM$ in \Equation{Eq:1040} are the ones before shifting the initial-state variables.
We consider here the $\inlbl$-case, so $\Qin\to\Qin+\xP_r\pP+p_{r\theperp}$ and we need to transform $\xP\to\xP+\xP_r$ to get
%
\begin{align}
\LXYZ{\Icol}{\inlbl r}
&=
\frac{1}{\piep\mu^\vep}
\,\frac{2}{\sTot}
\int d^{4+\vep}p_r\,\delta_+(p_r^2)
\,\SXYZ{\Icol}{\inlbl r}(\xP_r,\xM_r)
\,\ELshiftIin{p_r}
\,\theta\big(\xP_r<1-\xP\big)\,\theta\big(\xM_r<1-\xM\big)
~,
\end{align}
%
with
%
\begin{align}
\SXYZ{\Icol}{\inlbl r}(\xP_r,\xM_r)
&= 
\theta\big(\xM_r<\xiP\xP_r\big)
\,\frac{-1}{\xM_r(\xP+\xP_r)}
\,\EuScript{Q}_{\inlbl r}\bigg(\frac{-\xP_r}{\xP+\xP_r}\bigg)
~,
\end{align}
%
and now
%
\begin{equation}
\ELshiftIin{p_r}
=
\left\{
\begin{array}{ll}
  \EL{\big(\Qin+\xP_r\pP+p_{r\theperp};\pSetN\big)}/\EL{\big(\Qin;\pSetN\big)}
       & \textrm{conservative approach} \\
  \EL{\big(\Qin+\xP_r\pP;\pSetN\big)}/\EL{\big(\Qin;\pSetN\big)}
  =
  \ell_{\inlbl}(\xP+\xP_r)
       & \textrm{progressive approach}\\
\end{array} 
\right. 
\end{equation}
%
with
%
\begin{equation}
\ell_{\inlbl}(y)
=
\frac{\EL{\big(y\pP+\xM\pM+\kperp;\pSetN\big)}}
     {\EL{\big(\xP\pP+\xM\pM+\kperp;\pSetN\big)}}
~.
\label{Eq:1359}
\end{equation}
%
%
The $\inbar$-case is completely analogous, and we keep $\EuScript{Q}_{\inlbl}$ general without any assumptions.
It is useful to already write the integral over $p_r$ explicitly in terms of the Sudakov variables
%
\begin{equation}
\LXYZ{\Icol}{\inlbl r}
=
\frac{1}{\piep\mu^\vep}
\int_0^{1-\xP}d\xP_r\int_0^{1-\xM}d\xM_r
\,\SXYZ{\Icol,r}{\inlbl}(\xP_r,\xM_r)
\int d^{2+\vep}p_{r\theperp}\delta_+\big(\sTot\xP_r\xM_r-|p_{r\theperp}|^2\big)
\,\ELshiftIin{p_r}
~.
\end{equation}
%
In principle we could define $\LXYZ{\Icoldiv}{\inlbl r}$ such that $\LXYZ{\Icolfin}{\inlbl r}$ vanishes in the progressive case.
However, we prefer to keep $\LXYZ{\Icoldiv}{\inlbl r}$ simpler at the cost of non-vanishing $\LXYZ{\Icolfin}{\inlbl r}$, by removing the condition $\xM_r<1-\xM$ for the former.
Using
%
\begin{equation}
\int d^{2+\vep}p_{r\theperp}\delta_+\big(\sTot\xP_r\xM_r-|p_{r\theperp}|^2\big)
=
\piep\big(\sTot\xM_r\xP_r\big)^{\vep/2}
\end{equation}
%
we define
%
\begin{align}
\LXYZ{\Icolfin}{\inlbl r}
&=
\int_0^1d\xP_r\int_0^1d\xM_r
\,\SXYZ{\Icol}{\inlbl r}(\xP_r,\xM_r)\,\theta(\xP_r<1-\xP)
\\&\hspace{8ex}\times
\bigg[
\int\frac{d^{2}p_{r\theperp}}{\pi}\delta_+\big(\sTot\xP_r\xM_r-|p_{r\theperp}|^2\big)
  \ELshiftIin{p_r}\,\theta(\xM_r<1-\xM)
-
\ell_{\inlbl}(\xP+\xP_r)
\bigg]
~,\notag
\end{align}
%
and
%
\begin{align}
\LXYZ{\Icoldiv}{\inlbl r}
&=
\bigg(\frac{\sTot}{\mu^2}\bigg)^{\!\vep/2}
\int_0^1d\xP_r\xP_r^{\vep/2}\int_0^1d\xM_r\xM_r^{\vep/2}
\,\SXYZ{\Icol}{\inlbl r}(\xP_r,\xM_r)
\,\ell_{\inlbl}(\xP+\xP_r)\,\theta(\xP_r<1-\xP)
\\&\hspace{0ex}=
\frac{2}{\vep}
\bigg(\frac{\sTot\xP^2\xiP}{\mu^2}\bigg)^{\!\vep/2}
\int_0^1\frac{d\xP_r}{\xP}\bigg(\frac{\xP_r}{\xP}\bigg)^{\!\vep}
\,\frac{-\xP}{\xP+\xP_r}
\,\EuScript{Q}_{\inlbl r}\bigg(\frac{-\xP_r}{\xP+\xP_r}\bigg)
\,\ell_{\inlbl}(\xP+\xP_r)\,\theta(\xP_r<1-\xP)
\notag~.
\end{align}
%
The following variable substitutions now make sense
%
\begin{equation}
\zP = \frac{\xP}{\xP+\xP_r}
\;\;\Leftrightarrow\;\;
\xP_r = \frac{1-\zP}{\zP}\,\xP
\;\;,\;\;
d\xP_r = \xP\,\frac{d\zP}{\zP^2}
~,
\end{equation}
%
leading to
%
\begin{align}
\LXYZ{\Icoldiv}{\inlbl r}
&=
\frac{2}{\vep}
\bigg(\frac{\sTot\xP^2\xiP}{\mu^2}\bigg)^{\!\vep/2}
\int_0^1d\zP\bigg(\frac{1-\zP}{\zP}\bigg)^{\!\vep}
\,\EuScript{P}_{\inlbl r}(\zP)
\,\frac{\ell_{\inlbl}(\xP/\zP)}{z^2}\,\theta(\zP>\xP)
~,
\end{align}
%
where we used the fact that
%
\begin{equation}
{-}z\EuScript{Q}_{\inlbl r}(\zP-1) = \EuScript{P}_{\inlbl r}(z)
~,
\end{equation}
%
see \Appendix{App:splittingfunctions}~.
The possible flip $gq\leftrightarrow qq$ for the $\inbar$ case is understood as explained there.
A factor $1/z^2$ appears instead of the $1/z$ one might have expected because $\EL{}$ includes the flux factor.

The splitting function has a term $2C_{\inlbl r}/(1-z)$, causing a soft singularity.
(For a $\inbar$-case without such a term we just set $C_{\inbar r}=0$ in the following.)
We can isolate the singularity via
%
\begin{align}
\LXYZ{\Icoldiv}{\inlbl r}
&=
\frac{2}{\vep}
\bigg(\frac{\sTot\xP^2\xiP}{\mu^2}\bigg)^{\!\vep/2}
\Bigg\{
\frac{2C_{\inlbl r}}{\vep}
+
\int_0^1d\zP(1-z)^{\vep}\bigg[
\,\EuScript{P}_{\inlbl r}(\zP)
\,\zP^{-\vep}
\,\frac{\ell_{\inlbl}(\xP/\zP)}{z^2}\,\theta(\zP>\xP)
-\frac{2C_{\inlbl r}}{1-z}\bigg]
\Bigg\}
\notag\\&\hspace{0ex}=
\bigg(\frac{\sTot\xP^2\xiP}{\mu^2}\bigg)^{\!\vep/2}
\Bigg\{
\frac{4C_{\inlbl r}}{\vep^2}
+\frac{2}{\vep}\int_0^1d\zP\big[1+\vep\ln(1-z)\big]
\,\EuScript{P}_{\inlbl r}^{\mathrm{reg}}(\zP)
\,\frac{\ell_{\inlbl}(\xP/\zP)}{z^2}\,\theta(\zP>\xP)
\\&\hspace{30ex}-
2\int_0^1d\zP
\,\ln(z)\,\EuScript{P}_{\inlbl r}(\zP)
\,\frac{\ell_{\inlbl}(\xP/\zP)}{z^2}\,\theta(\zP>\xP)
+\Ord(\vep)
\Bigg\}
~,
\notag
\end{align}
%
where
%
\begin{equation}
\EuScript{P}_{\inlbl r}^{\mathrm{reg}}(\zP)
=
\EuScript{P}_{\inlbl r}(\zP)
 - \frac{2C_{\inlbl r}}{1-z}
+\frac{2C_{\inlbl r}}{[1-z]_+}
~,
\label{Eq:1522}
\end{equation}
%
and where the plus-distribution only acts towards the right:
%
\begin{equation}
\int_0^1dz\,f(z)\,\frac{1}{[1-z]_+}\,g(z) = 
\int_0^1dz\,f(z)\,\frac{1}{1-z}\,[g(z)-g(1)] 
~.
\end{equation}

\subsection{Final results for the divergent parts}
We calculate the divergent contributions $\LXYZ{X,Y,\mathrm{div}}{Z}(\vepv)$ in \Appendix{App:divergent} (there in terms of $\vep=-2\vepv$), and bring the expressions for $\LXYZ{X,Y,\mathrm{fin}}{Z}$ to a suitable form for numerical integration in \Appendix{App:finite}.
Here, we present the final results for the former.
The final-state collinear ones are
%
\begin{align}
\LXYZ{\Fcoldiv}{ir}(\vepv)
&=
\bigg(\frac{\mu^2}{4E_i^2}\bigg)^{\!\vepv}
\bigg[{-}\frac{1}{\vepv}+ \ln\zeta_0+\vepv\Big(\mathrm{Li}_2(\zeta_0)-\frac{1}{2}\ln^2\zeta_0\Big)+\Ord\big(\vepv^2\big)\bigg]\,I_{ir}(\vepv)
~,
\end{align}
%
where $I_{ir}(\vepv) = \int_0^{1/2}\hspace{0ex}dx\,x^{-2\vepv}\,\EuScript{P}_{ir}(1-x)$, that is 
%
\begin{align}
I_{i\to g,r\to g}(\vepv) &=
C_g\bigg[
    {-}\frac{1}{\vepv} - \frac{11}{6} 
     - \vepv\bigg(\frac{11\ln2}{2}-\frac{\pi^2}{3}+\frac{137}{36}\bigg)
+\Ord\big(\vepv^2\big)
\bigg]
~,
\\
I_{i\to q,r\to g}(\vepv) &=
C_q\bigg[
    {-}\frac{1}{\vepv} - 2\ln2 - \frac{7}{8}
     - \vepv\bigg(2\ln^22+\frac{7\ln2}{4} + 2\bigg)
+\Ord\big(\vepv^2\big)
\bigg]
\label{Eq:919}~,
\\
I_{i\to g, r\to q}(\vepv) &=
C_q\bigg[
      2\ln2 - \frac{5}{8}
     + \vepv\bigg(2\ln^22-\frac{5\ln2}{4} +\frac{\pi^2}{3} - \frac{3}{2}\bigg)
+\Ord\big(\vepv^2\big)
\bigg]
\label{Eq:927}~,
\\
I_{i \to q,r\to q}(\vepv) &=
T_R\bigg[
      \frac{1}{3}
     + \vepv\bigg(\frac{2\ln2}{3} +\frac{23}{36} \bigg)
+\Ord\big(\vepv^2\big)
\bigg]
\label{Eq:2174}
~.
\end{align}
%
The soft and soft-collinear contributions are
%
\begin{align}
\LXYZ{\Fsoftdiv}{ib}(\vepv)
&=
\bigg(\frac{\mu^2}{2E_0^2\,\lop{n_i}{n_b}}\bigg)^{\!\vepv}
\bigg[{-}\frac{1}{\vepv^2}-\frac{1}{\vepv}\frac{2E_0}{E_i} + \frac{\pi^2}{6}-\frac{4E_0}{E_i}+\Ord\big(\vepv\big)\bigg]
~.
\end{align}
%
\begin{align}
\LXYZ{\Fsoftcoldiv}{i}(\vepv)
&=
C_i
\bigg(\frac{\mu^2}{4E_0^2}\bigg)^{\!\vepv}
\bigg[{-}\frac{1}{\vepv^2}
      +\frac{1}{\vepv}\bigg(\ln\zeta_0-\frac{2E_0}{E_i}\bigg)
\\&\hspace{16ex}
      +\frac{2E_0}{E_i}\ln\zeta_0 + \mathrm{Li}_2(\zeta_0)-\frac{1}{2}\ln^2\zeta_0
      -\frac{4E_0}{E_i}
      +\Ord\big(\vepv\big)
\bigg]
\notag~.
\end{align}
%
\begin{align}
\LXYZ{\Isoftdiv}{\inlbl b}(\vepv)
&=
\bigg(\frac{\mu^2}{2E_0^2\,\lop{n_{\inlbl}}{n_b}}\bigg)^{\!\vepv}
\bigg[{-}\frac{1}{\vepv^2} + \frac{\pi^2}{6}+\Ord\big(\vepv\big)\bigg]
~.
\end{align}
%
%
\begin{align}
\LXYZ{\Isoftcoldiv}{\inlbl}(\vepv)
&=
C_{\inlbl}
\bigg(\frac{\mu^2E}{4E_0^2\bar{E}\xiP}\bigg)^{\!\vepv}
\bigg[{-}\frac{1}{\vepv^2}-2\mathrm{Li}_2\bigg({-}\frac{\bar{E}\xiP}{E}\bigg)+\Ord\big(\vepv\big)\bigg]
~.
\end{align}
The initial-state collinear contributions are
%
\begin{align}
\LXYZ{\Icoldiv}{\inlbl r}(\vepv)
&=
\bigg(\frac{\mu^2E}{4E^2\xP^2\bar{E}\xiP}\bigg)^{\!\vepv}
\Bigg\{
\frac{C_{\inlbl r}}{\vepv^2}
-\frac{1}{\vepv}\int_0^1d\zP
\,\EuScript{P}_{\inlbl r}^{\mathrm{reg}}(\zP)
\,\frac{\ell_{\inlbl}(\xP/\zP)}{z^2}\,\theta(\zP>\xP)
\\&\hspace{-3ex}
+2\int_0^1d\zP\Big[\ln(1-z)
\,\EuScript{P}_{\inlbl r}^{\mathrm{reg}}(\zP)
-\ln(z)\,\EuScript{P}_{\inlbl r}(\zP)
-\srac{1}{2}\EuScript{P}_{\inlbl r}^{(1)}(\zP)
\Big]\frac{\ell_{\inlbl}(\xP/\zP)}{z^2}\,\theta(\zP>\xP)
+\Ord(\vepv)
\Bigg\}
~,
\notag
\end{align}
%
where $\EuScript{P}_{\inlbl r}^{\mathrm{reg}}(\zP)$ is defined in \Equation{Eq:1522}.
The splitting functions are evaluated at $\vepv=0$, and $\EuScript{P}_{\inlbl r}^{(1)}$ is the $\Ord(\vepv)$ coefficient.
The results for the $\inbar$ case are completely analogous with $\xP\leftrightarrow\xM$, $\pP\leftrightarrow\pM$, and $E\leftrightarrow\bar{E}$.
If the radiation is a quark or antiquark, we simply have $C_{\inbar r}=0$ in the formula above.
The function $\ell_{\inlbl}$ was defined in \Equation{Eq:1359}.

The expressions above belong to the integrals of the subtraction term with label $r$ in \Equation{Eq:229}. 
If $r$ does not refer to a gluon, then the terms labelled `$\soft$' and `$\soco$' vanish.
Let us consider the terms for which $r$ refers to a gluon ($r\to g$), and let us abbreviate
%
\begin{equation}
\MtreeCorNorm{a}{b} = \MtreeCor{a}{b}/\Mtree{}
~.
\end{equation}
%
Using color conservation, it is straighforeward to check that
%
\begin{align}
\LXYZ{\Fcoldiv}{i,r\to g}(\vepv)
+\LXYZ{\Fsoftcoldiv}{i}(\vepv)
+\sum_{b\neq i}\LXYZ{\Fsoftdiv}{ib}(\vepv)\MtreeCorNorm{i}{b}
&=
\mathrm{Soft}_{i}(\vepv) 
+\frac{\gamma^{(g)}_i}{\vepv}
+\Ord\big(\vepv^0\big)
\label{Eq:1106}
\\
\LXYZ{\Icoldiv}{\inlbl r\to g}(\vepv)
+\LXYZ{\Isoftcoldiv}{\inlbl}(\vepv)
+\sum_{b\neq\inlbl}\LXYZ{\Isoftdiv}{\inlbl b}(\vepv)\MtreeCorNorm{\inlbl}{b}
&=
\mathrm{Soft}_{\inlbl}(\vepv) 
\notag\\&\hspace{-12ex}
-\frac{1}{\vepv}\int_0^1d\zP
\,\EuScript{P}_{\inlbl r\to g}^{\mathrm{reg}}(\zP)
\,\frac{\ell_{\inlbl}(\xP/\zP)}{z^2}\,\theta(\zP>\xP)
+\Ord\big(\vepv^0\big)
\label{Eq:1098}
\end{align}
%
where
%
\begin{equation}
\mathrm{Soft}_{a}(\vepv) = \frac{C_{a}}{\vepv^2}
-\frac{C_a}{\vepv}\ln\bigg(\frac{\mu^2}{E_a^2}\bigg)
+\frac{1}{\vepv}\sum_{b\neq a}\ln\bigg(\frac{1}{2\lop{n_{a}}{n_b}}\bigg)
             \MtreeCorNorm{a}{b}
\end{equation}
%
and
%
\begin{equation}
\gamma^{(g)}_i = C_i\times
\left\{
\begin{array}{ll}
11/6 &\quad\textrm{if $i$ refers to a gluon,}\\
2\ln2 + 7/8 &\quad\textrm{if $i$ refers to a quark or antiquark.}\\
\end{array} 
\right. 
\label{Eq:1134}
\end{equation}
%
Clearly, the divergences are independent of $E_0,\zeta_0,\xi_0$.
The sum in $\mathrm{Soft}_{a}(\vepv)$ is only over $b$.
Summing also over $a$ and applying color conservation again, we see that $\sum_a\mathrm{Soft}_{a}(\vepv)$ gives the well-known universal expression for the soft and soft-collinear divergences (\cf~\cite{Somogyi:2006cz}):
using
%
\begin{equation}
\sum_{a,b;b\neq a}\ln\bigg(\frac{\mu^2}{2\lop{p_{a}}{p_b}}\bigg)\MtreeCorNorm{a}{b}
=
-2\sum_{a}C_a\ln\bigg(\frac{\mu}{E_a}\bigg)
+
\sum_{a,b;b\neq a}\ln\bigg(\frac{1}{2\lop{n_{a}}{n_b}}\bigg)\MtreeCorNorm{a}{b}
\end{equation}
%
we find
%
\begin{align}
\sum_{a}\mathrm{Soft}_{a}(\vepv) 
&= \sum_{a}\frac{C_{a}}{\vepv^2}
+\frac{1}{\vepv}\sum_{a,b;b\neq a}\ln\bigg(\frac{\mu^2}{2\lop{p_{a}}{p_b}}\bigg)
             \MtreeCorNorm{a}{b}
\notag\\&\hspace{0ex}=
\sum_{a,b}\bigg(\frac{\mu^2}{2\lop{p_{a}}{p_b}}\bigg)^{\!\vepv}
                  \frac{\MtreeCorNorm{a}{b}}{\vepv^2}
+\Ord\big(\vepv^0\big)
\quad\textrm{with}\quad \MtreeCorNorm{a}{a} = C_a
\label{Eq:2379}
~.
\end{align}
%
The collinear left-over in \Equation{Eq:1098} is also exactly as expected, and is the one that needs to be cancelled against a collinear counter term as prescribed by factorization.
This is well-known for the on-shell initial state, the one with $\inbar$ instead of $\inlbl$, and in literature often identified as part of the ``$P$-operator'', but shown in~\cite{vanHameren:2022mtk} to be also present for the space-like initial state.
There is a $1/z^2$ instead of the usual $1/z$ because $\ell_{\inlbl}$ includes the flux factor.

Except the initial-state collinear left-over, the divergences should cancel against those in the virtual contribution at NLO coming from the one-loop amplitudes.
The virtual contribution involves the same phase space integral as the Born contributions, but has $2\mathrm{Re}\big\{\EuScript{M}^\dagger\EuScript{M}^{\mathrm{one-loop}}(\vepv)\big\}$ instead of the Born-level matrix element $|\EuScript{M}|^2$.
It was argued in \cite{vanHameren:2022mtk} that the cancellation should happen exactly the same way for the case with a space-like initial-state as for the completely on-shell case.
More precisely, the divergences in the one-loop amplitudes that do not fit in the universal formula (as given for example in \cite{Catani:1998bh}) where identified in \cite{vanHameren:2022mtk}, and confirmed in~\cite{Blanco:2022iai}, and found to be universal themselves, so they can be treated within a renormalization prescription.
In the following, we show that the divergences from the real radiation we found here indeed match the one-loop formula as in~\cite{Catani:1998bh}.

The formulas for the divergent contributions we found above are the result of concentrating on a given radiative process.
To cancel all divergences, one must include all sub processes contributing to a given jet process. 
The formula for the total divergent soft contribution \Equation{Eq:2379} however already matches the universal formula for the soft contribution in one-loop amplitudes.
This comes out nicely, because removing a soft gluon from the radiative process exactly results in the Born process for which both the one-loop amplitude and the radiative process form the NLO correction.
Including all collinear divergences for which at least one gluon is involved is easily achieved by adding the contribution when a radiative quark becomes collinear to a gluon from \Equation{Eq:927} to the $\gamma^{(g)}_q$ of \Equation{Eq:1134}, leading to
%
\begin{equation}
\gamma_g=\gamma^{(g)}_g = \srac{11}{6}C_g
\quad,\quad
\gamma_q=\gamma^{(g)}_q + \gamma^{(q)}_g = \srac{3}{2}C_q
~.
\end{equation}
%
We see that also $\gamma_q$ now matches the expression for the one-loop amplitude.

In order to get the fermion-loop contribution for $\gamma_g$ included, one must go beyond concentrating on a single Born process.
Then one sees that a radiative process for which final-state quark-antiquark pair becomes collinear, there is a collinear divergence of $-T_R/(3\vepv)$ (from \Equation{Eq:2174}) to the Born process for which that pair is replaced by a gluon.
The same for a final-state quark and initial-state quark.
Then there is a factor $2$ for the contribution for which the quark and anti-quark reverse roles, and a factor counting the number of quark families included, leading to $\gamma_g=\frac{11}{6}C_g-\frac{2}{3}T_Rn_f$ from the one-loop formula.

\section{Numerical checks}
We implemented the presented scheme using the framework of \KaTie.
In particular, we augmented the phase space generator {\sc Kaleu} (\cite{vanHameren:2010gg}) with the channels matching the subtraction terms within the adaptive multi-channel method~\cite{Kleiss:1994qy}, in order to achieve convergent Monte Carlo estimates of the phase space integrals.
In this section, we present some numerical results in support of the viability of the proposed scheme.
We show that it indeed leads to convergent subtracted-real integrals, and that their combination with the $\Ord\big(\vepv^0\big)$ part of the integrated subtraction terms is independent of the parameters $E_0$ and $\zeta_0$ (\Equation{Eq:612} and \Equation{Eq:613}).
We link $\xi_0$ of \Equation{Eq:1040} to $\zeta_0$ via \Equation{Eq:4419}.
The phase space and renormalization/factorization scale are defined by
%
\begin{gather}
E_{\mathrm{cm}}=14\mathrm{TeV}
\;,\;\;
\textrm{anti-$k_T$ algorithm with $R=0.4$}
\;,\;\;
p_{T,\mathrm{jet}}>50\mathrm{GeV}
\;,\;\;
|y_{\mathrm{jet}}|<4
\;,\notag\\
\mu_{R}=\mu_{F}=\frac{1}{2}\sum_{i}p_{T,i}
~,
\label{Eq:2464}
\end{gather}
where the sum is over all final-state partons.

\begin{figure}
\begin{center}
\epsfig{figure=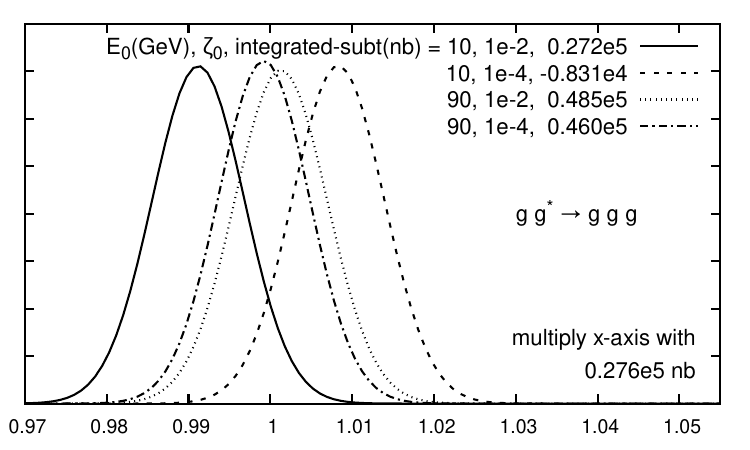,width=0.495\linewidth}\hfill
\epsfig{figure=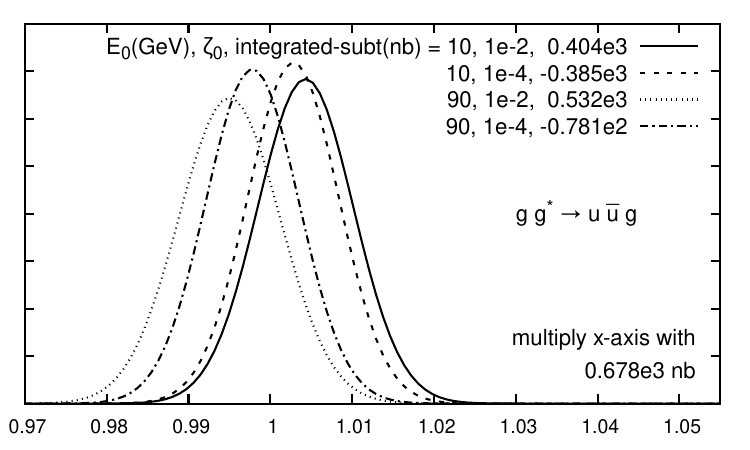,width=0.495\linewidth}
\epsfig{figure=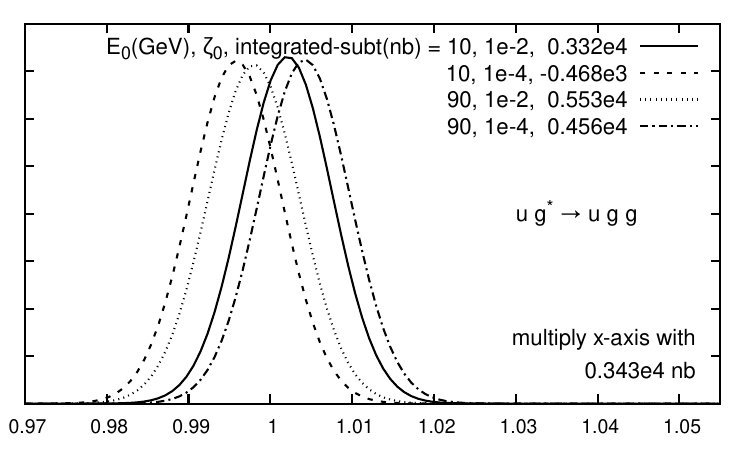,width=0.495\linewidth}\hfill
\epsfig{figure=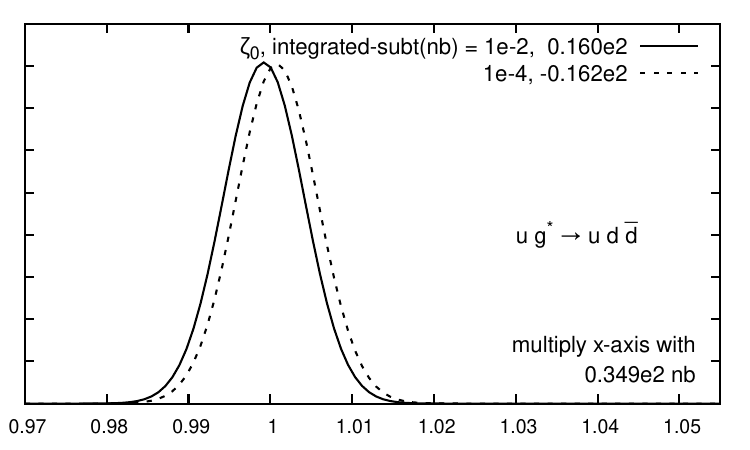,width=0.495\linewidth}
\caption{Statistical distributions of the Monte Carlo estimate of the $\Ord\big(\vepv^0\big)$ part of the real-radiation contribution to the NLO cross section for dijet production, for different values of the parameters $E_0,\zeta_0$. The essence of the plots is that the distributions have finite width, proving convergence, and that they overlap, proving independence of the parameters $E_0,\zeta_0$. The values of the integrated subtraction terms (which are precise below the per mille level) are written separately, to indicate that the same final results indeed come from different individual terms. The phase space and renormalization/factorization scale are defined in \Equation{Eq:2464}. 
} \label{Fig:001}
\end{center}
\end{figure}
In \Figure{Fig:001} we illustrate convergence and independence for a selection of partonic radiative processes relevant to dijet production.
The parton branching PDFs {\tt PB-NLO-HERAI+II-2018\-set2}~\cite{BermudezMartinez:2018fsv}, taken from {\sc TMDlib}~\cite{Abdulov:2021ivr}, were employd for the space-like side, and {\tt CT18nlo}~\cite{Hou:2019qau}, taken from {\sc LHAPDF}~\cite{Buckley:2014ana}, for the on-shell side.

\begin{figure}
\begin{center}
\epsfig{figure=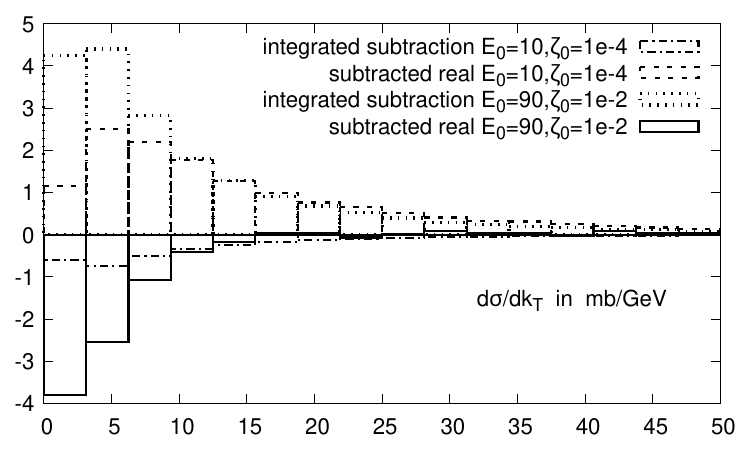,width=0.495\linewidth}\hfill
\epsfig{figure=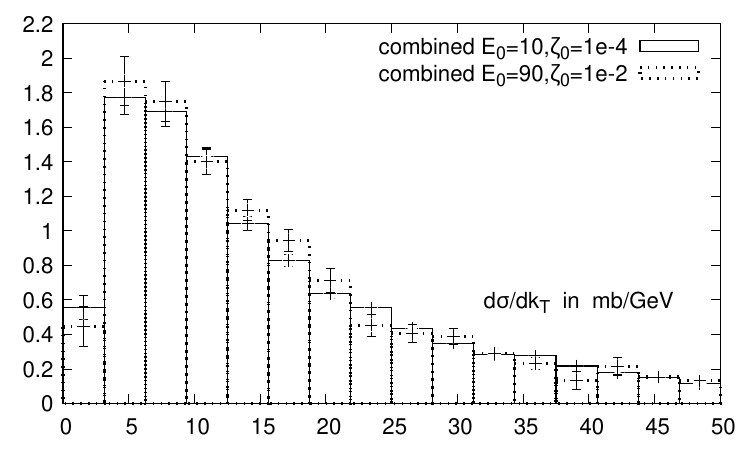,width=0.495\linewidth}
\epsfig{figure=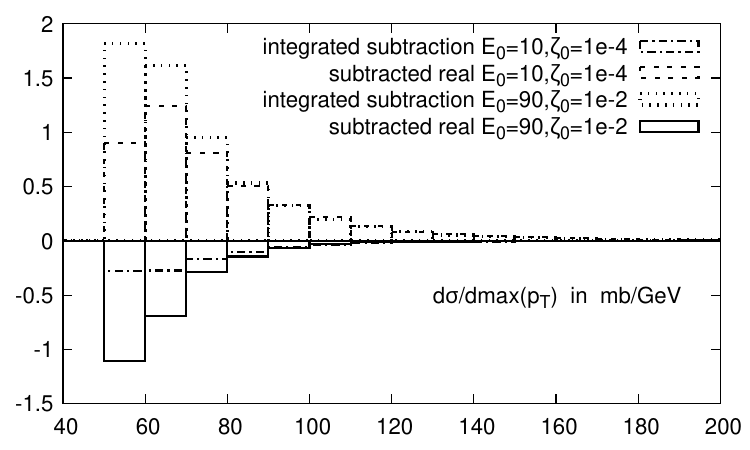,width=0.495\linewidth}\hfill
\epsfig{figure=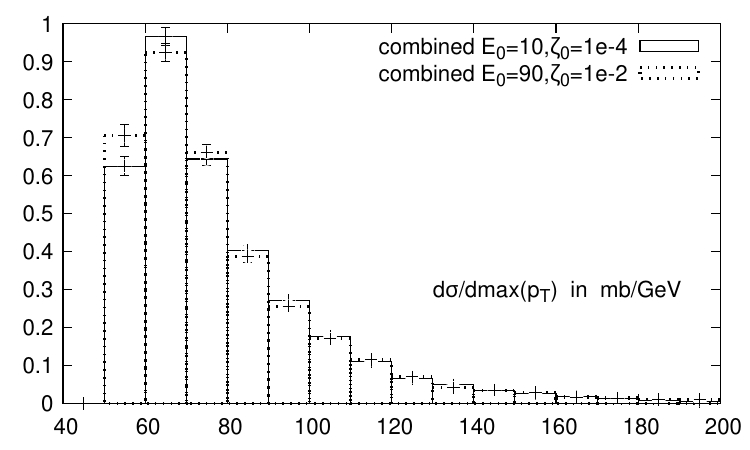,width=0.495\linewidth}
\caption{Cross sections differential in the final-state momentum imbalance (or initial-state) $k_T=|\kperp|$ (upper plots), and the $p_T$ of the hardest jet (lower plots) for dijet production. On the left are the individual plots for the subtracted-real integral and the $\Ord\big(\vepv^0\big)$ part of the integrated subtraction terms, for two values of the parameters $E_0,\zeta_0$. Notice that the individual contributions flip sign going from one parameter set to the other. On the right are the sum of subtracted-real and integrated subtraction, clearly independent of the parameters within statistical accuracy. The processes included form a representative selection, and are given by
$g g^\star\to3g$,
$g g^\star\to u\bar{u}g$,
$u g^\star\to u g g$,
$u g^\star\to u d \bar{d}$,
$u g^\star\to u u \bar{u}$,
and those with $u\leftrightarrow d$.
The phase space and renormalization/factorization scale are defined in \Equation{Eq:2464}.
}
\label{Fig:002}
\end{center}
\end{figure}
While \Figure{Fig:001} present results for the full cross section for separate processes, \Figure{Fig:002} presents results for differential cross sections for a sum of processes, again relevant to dijet production.
The results clearly indicate independence of the parameters $E_0,\zeta_0$ despite the rather substantial cancellation between the individual contributions.
The computation were performed on a single modern laptop with $12$ cores.

\section{On-shell limit}
The on-shell limit $|\kperp|\to0$ for tree-level matrix elements with a space-like gluon with momentum $k_{\inlbl}^\mu$ (\Equation{Eq:454}) is given by
%
\begin{equation}
\big|\EuScript{M}(\kperp)\big|^2
 \;\overset{|\kperp|\to0}{-\hspace{-0.2ex}-\hspace{-1.0ex}\longrightarrow}\;
\EuScript{M}^*_\mu(0)\,\frac{\kperp^\mu\kperp^\nu}{|\kperp|^2}\,\EuScript{M}_\nu(0)
~.
\end{equation}
%
Here, we only make the dependence on $\kperp$ explicit as argument of $\EuScript{M}$, and denote by $\EuScript{M}_\nu(0)$ the on-shell amplitude with the polarization vector for the, now on-shell, gluon stripped off.
The on-shell limit is ``smooth'' only including integration over the remaining azimuthal angle dependence, which is turned into a sum over helicities
%
\begin{equation}
\int_0^{2\pi}\frac{d\varphi_{\theperp}}{2\pi}\,\big|\EuScript{M}(\kperp)\big|^2
 \;\overset{|\kperp|\to0}{-\hspace{-0.2ex}-\hspace{-1.0ex}\longrightarrow}\;
\big|\EuScript{M}(0)\big|^2
~.
\end{equation}
%
For tree-level cross sections, this limit is also smooth, and can be imagined to be established with the help of a special $\kperp$-dependent PDF that has a free parameter $\beta$ such that
%
\begin{equation}
F(\kperp,x;\beta)
 \;\overset{\beta\to0}{-\hspace{-1.0ex}\longrightarrow}\;
\delta(\kperp^2)\,f(x)
~,
\end{equation}
%
so
%
\begin{equation}
\int\frac{d^2\kperp}{\pi}\,F(\kperp,x;\beta)\,\big|\EuScript{M}(\kperp)\big|^2
 \;\overset{\beta\to0}{-\hspace{-1.0ex}\longrightarrow}\;
f(x)\,\big|\EuScript{M}(0)\big|^2
~.
\end{equation}
%

Beyond tree-level, the smooth limits do not hold anymore, as explicated in~\cite{vanHameren:2022mtk}.
What we call real-radiation contribution in this write-up is not the whole NLO real contribution, but only the part that cannot be calculated analytically and that in~\cite{vanHameren:2022mtk} is referred to as the ``familiar'' real contribution.
Still, one could imagine to construct an artificial real integral that would lead to the on-shell case, or one could try to construct space-like subtraction terms to be applied to the one-shell real integral. 

Unfortunately, this cannot work, and what is worse, even in the $\kperp$-dependent case the subtraction method presented here strictly speaking fails at $|\kperp|=0$.
The reason is that at $|\kperp|=0$ there is no {\em point-wise} cancellation between the radiative matrix element and the subtraction terms.
Let us denote the radiative matrix element by $\big|\EuScript{M}(\kperp,\rperp)\big|^2$, so with $2$ arguments, where $\rperp$ refers to the transverse part of the recoil of the radiation, and by $\big|\EuScript{M}(\kperp-\rperp)\big|^2$, with only $1$ argument, the matrix element appearing in a subtraction term.
The latter has one final-state parton fewer, and the radiative recoil subtracted from the initial state.
For the situation in which $|\kperp|\to0$ before $|\rperp|\to0$, we schematically have
\newcommand{\Singular}{\mathrm{S}{\scriptstyle\mathrm{ingular}}}
\begin{equation}
\big|\EuScript{M}(\kperp,\rperp)\big|^2
 \;\overset{|\kperp|\to0}{-\hspace{-1.0ex}\longrightarrow}\;
\EuScript{M}^*_\mu(0,\rperp)\,\frac{\kperp^\mu\kperp^\nu}{|\kperp|^2}\,\EuScript{M}_\nu(0,\rperp)
 \;\overset{|\rperp|\to0}{-\hspace{-1.0ex}\longrightarrow}\;
\Singular\times\EuScript{M}^*_\mu(0)\,\frac{\kperp^\mu\kperp^\nu}{|\kperp|^2}\,\EuScript{M}_\nu(0)
\end{equation}
%
while
%
%
\begin{equation}
\Singular\times\big|\EuScript{M}(\kperp-\rperp)\big|^2
 \;\overset{|\kperp|\to0}{-\hspace{-1.0ex}\longrightarrow}\;
\Singular\times\big|\EuScript{M}(-\rperp)\big|^2
 \;\overset{|\rperp|\to0}{-\hspace{-1.0ex}\longrightarrow}\;
\Singular\times\EuScript{M}^*_\mu(0)\,\frac{\rperp^\mu\rperp^\nu}{|\rperp|^2}\,\EuScript{M}_\nu(0)
\end{equation}
%
and we see that the two cases do not match.
Only after integration over the remnant azimuthal angle dependence of both $\kperp$ and $\rperp$, the cases match.
So while the integrated subtraction terms correctly represent the divergences, at $|\kperp|=0$ the terms fail at the task of point-wise cancellation of the singularities in the subtracted real integral.

The reason why the subtraction method presented in this write-up still works is that the measure of phase space where this problem appears vanishes.
Both space-like matrix elements and $\kperp$-dependent PDFs are in practice defined such that they are finite for $|\kperp|\to0$, while the $2$-dimensional integration over $\kperp$ provides a factor $|\kperp|$, so the differential cross section vanishes.
\begin{figure}
\begin{center}
\epsfig{figure=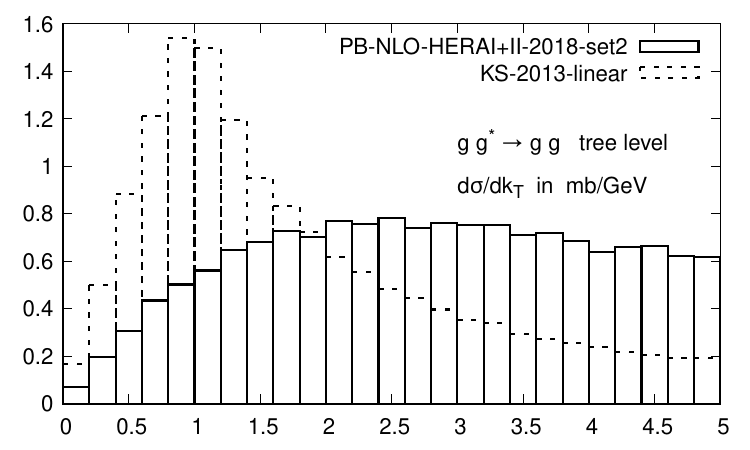,width=0.495\linewidth}\hfill
\epsfig{figure=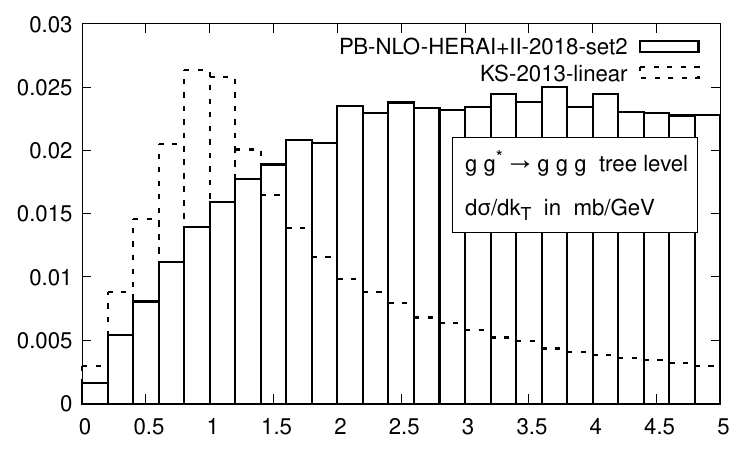,width=0.495\linewidth}
\epsfig{figure=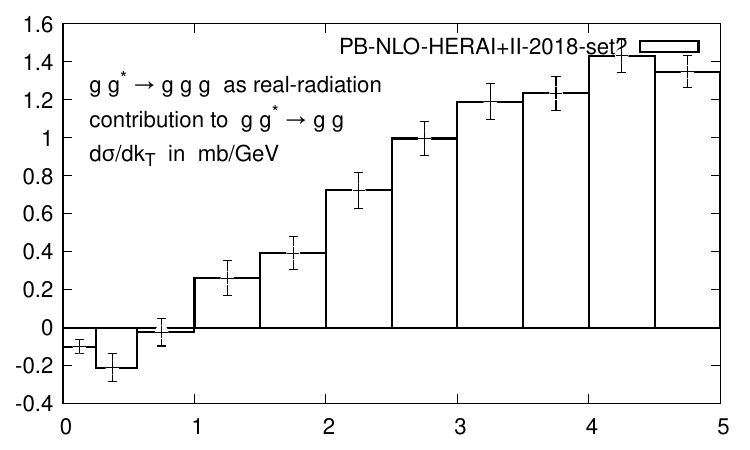,width=0.495\linewidth}
\epsfig{figure=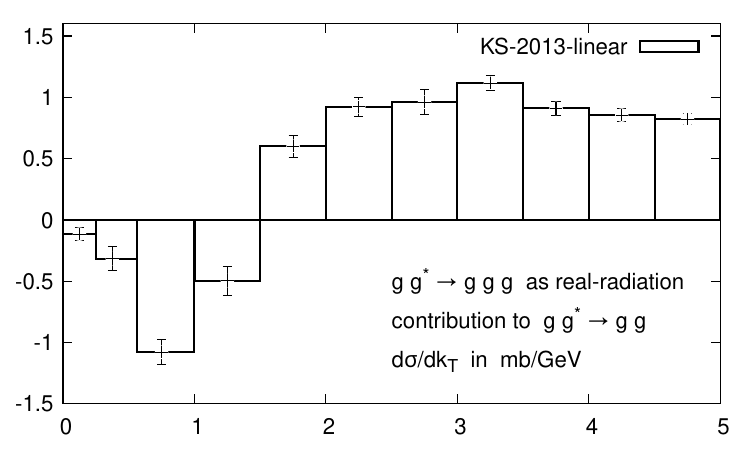,width=0.495\linewidth}
\caption{Cross sections differential in the final-state momentum imbalance (or initial-state) $k_T=|\kperp|$. The essence of the plots is that the distributions vanish for $k_T\to0$.
The upper plots show LO distributions, for $g g^\star\to g g$ and $g g^\star\to g g g$, and for two different $k_T$-dependent PDFs.
The lower row shows the $\Ord\big(\vepv^0\big)$-part of the real-radiation contribution to $g g^\star\to g g$ from the $g g^\star\to g g g$ channel, for both $k_T$-dependent PDFs.
The phase space and renormalization/factorization scale are defined in \Equation{Eq:2464}.
The $k_T$-dependent PDF {\tt KS-2013-linear} is set to zero beyond its range of validity $x<0.01$.} \label{Fig:003}
\end{center}
\end{figure}
To illustrate this, we present these differential cross sections for $g g^\star\to g g$ and $g g^\star\to g g g$ at tree level in the upper row of \Figure{Fig:003}.
We present results for both $k_T$-dependent PDFs {\tt PB-NLO-HERAI+II-2018\-set2} and {\tt KS-2013-linear}~\cite{Kutak:2012rf}.
While LO distribution of the latter does exhibit the, in this discussion {\em dangerous}, growth with $k_T\to0$ associated with linear BFKL evolution, it eventually still vanishes.
The lower row shows the $\Ord\big(\vepv^0\big)$-part of the real-radiation contribution to $g g^\star\to g g$ from the $g g^\star\to g g g$ channel, for both $k_T$-dependent PDFs.
We remind the reader that while the full real-radiation contribution must be positive, the $\Ord\big(\vepv^0\big)$-part on its own does not have to be positive.

\section{Conclusion}
We presented a subtraction scheme for the calculation of the real-radiation contribution in hybrid $k_T$-factorization.
It is different from existing schemes for collinear factorization in that the so-called momentum recoil is subtracted from the initial-state variables, leaving the final-state momenta untouched as much as possible.
We calculated the divergent part of the integrated subtraction terms, and show that they have the expected universal form.
The remaining finite part of these terms must be integrated numerically, and we presented the integrals in a form directly applicable for such a procedure.

We implemented the scheme and performed calculations for all processes relevant for $2$-jet production as NLO, and found that the subtracted real-radiation integrals indeed converge.
Furthermore, we found that the sum of these integrals and the integrated subtraction terms is independent of the parameters restricting the phase space on which the subtraction terms are set not to vanish.
Despite the fact that the phase space integration region for the integrated subtraction terms has $3$ extra dimensions because of the finite remainders, their computation time to reach a given precision is still only a fraction of the computation time for the subtracted real radiation integral.

%
%

\subsection*{Acknowledgments}
The authors would like to thank Rene Poncelet for useful discussions and remarks.
This work was supported by grant no.\ 2019/35/B/ST2/03531 of the Polish National Science Centre.
%


\begin{thebibliography}{10}

\bibitem{Somogyi:2006cz}
G.~Somogyi and Z.~Trocsanyi, {\it {A New subtraction scheme for computing QCD
  jet cross sections at next-to-leading order accuracy}},
  \href{http://xxx.lanl.gov/abs/hep-ph/0609041}{{\tt hep-ph/0609041}}.

\bibitem{Frixione:1995ms}
S.~Frixione, Z.~Kunszt, and A.~Signer, {\it {Three jet cross-sections to
  next-to-leading order}},  {\em Nucl.Phys.} {\bf B467} (1996) 399--442,
  [\href{http://xxx.lanl.gov/abs/hep-ph/9512328}{{\tt hep-ph/9512328}}].

\bibitem{Catani:1996vz}
S.~Catani and M.~Seymour, {\it {A General algorithm for calculating jet
  cross-sections in NLO QCD}},  {\em Nucl.Phys.} {\bf B485} (1997) 291--419,
  [\href{http://xxx.lanl.gov/abs/hep-ph/9605323}{{\tt hep-ph/9605323}}].

\bibitem{Somogyi:2009ri}
G.~Somogyi, {\it {Subtraction with hadronic initial states at NLO: An
  NNLO-compatible scheme}},  {\em JHEP} {\bf 05} (2009) 016,
  [\href{http://xxx.lanl.gov/abs/0903.1218}{{\tt 0903.1218}}].

\bibitem{Robens:2013wga}
T.~Robens, {\it {Nagy-Soper Subtraction: A Review}},  {\em Mod. Phys. Lett. A}
  {\bf 28} (2013) 1330020, [\href{http://xxx.lanl.gov/abs/1306.1946}{{\tt
  1306.1946}}].

\bibitem{Bevilacqua:2013iha}
G.~Bevilacqua, M.~Czakon, M.~Kubocz, and M.~Worek, {\it {Complete Nagy-Soper
  subtraction for next-to-leading order calculations in QCD}},  {\em JHEP} {\bf
  10} (2013) 204, [\href{http://xxx.lanl.gov/abs/1308.5605}{{\tt 1308.5605}}].

\bibitem{Fox:2023bma}
E.~Fox and N.~Glover,
{\it {Initial-final and initial-initial antenna functions for real radiation at next-to-leading order}},
{\em JHEP} {\bf 12} (2023) 171,
[\href{http://xxx.lanl.gov/abs/2308.10829}{{\tt 2308.10829}}].

\bibitem{Gehrmann-DeRidder:2005btv}
A.~Gehrmann-De~Ridder, T.~Gehrmann, and E.~W.~N. Glover, {\it {Antenna
  subtraction at NNLO}},  {\em JHEP} {\bf 09} (2005) 056,
  [\href{http://xxx.lanl.gov/abs/hep-ph/0505111}{{\tt hep-ph/0505111}}].

\bibitem{Somogyi:2006da}
G.~Somogyi, Z.~Trocsanyi, and V.~Del~Duca, {\it {A Subtraction scheme for
  computing QCD jet cross sections at NNLO: Regularization of doubly-real
  emissions}},  {\em JHEP} {\bf 01} (2007) 070,
  [\href{http://xxx.lanl.gov/abs/hep-ph/0609042}{{\tt hep-ph/0609042}}].

\bibitem{Czakon:2010td}
M.~Czakon, {\it {A novel subtraction scheme for double-real radiation at
  NNLO}},  {\em Phys. Lett. B} {\bf 693} (2010) 259--268,
  [\href{http://xxx.lanl.gov/abs/1005.0274}{{\tt 1005.0274}}].

\bibitem{Caola:2017dug}
F.~Caola, K.~Melnikov, and R.~R\"ontsch, {\it {Nested soft-collinear
  subtractions in NNLO QCD computations}},  {\em Eur. Phys. J. C} {\bf 77}
  (2017), no.~4 248, [\href{http://xxx.lanl.gov/abs/1702.01352}{{\tt
  1702.01352}}].

\bibitem{Magnea:2018hab}
L.~Magnea, E.~Maina, G.~Pelliccioli, C.~Signorile-Signorile, P.~Torrielli, and
  S.~Uccirati, {\it {Local analytic sector subtraction at NNLO}},  {\em JHEP}
  {\bf 12} (2018) 107, [\href{http://xxx.lanl.gov/abs/1806.09570}{{\tt
  1806.09570}}]. [Erratum: JHEP 06, 013 (2019)].

\bibitem{Collins:1991ty}
J.~C.~Collins and R.~K.~Ellis,
{\it {Heavy quark production in very high-energy hadron collisions}},
{\em Nucl. Phys. B} {bf 360} (1991), 3-30.

\bibitem{Catani:1990eg}
S.~Catani, M.~Ciafaloni and F.~Hautmann,
{\it {High-energy factorization and small x heavy flavor production}},
{\em Nucl. Phys. B} {\bf 366} (1991), 135-188.

\bibitem{Dumitru:2005gt}
A.~Dumitru, A.~Hayashigaki, and J.~Jalilian-Marian, {\it {The Color glass
  condensate and hadron production in the forward region}},  {\em Nucl. Phys.}
  {\bf A765} (2006) 464--482,
  [\href{http://xxx.lanl.gov/abs/hep-ph/0506308}{{\tt hep-ph/0506308}}].

\bibitem{Marquet:2007vb}
C.~Marquet, {\it {Forward inclusive dijet production and azimuthal correlations
  in p(A) collisions}},  {\em Nucl. Phys. A} {\bf 796} (2007) 41--60,
  [\href{http://xxx.lanl.gov/abs/0708.0231}{{\tt 0708.0231}}].

\bibitem{Deak:2009xt}
M.~Deak, F.~Hautmann, H.~Jung, and K.~Kutak, {\it {Forward Jet Production at
  the Large Hadron Collider}},  {\em JHEP} {\bf 0909} (2009) 121,
  [\href{http://xxx.lanl.gov/abs/0908.0538}{{\tt 0908.0538}}].

\bibitem{Nefedov:2019mrg}
M.~A. Nefedov, {\it {Computing one-loop corrections to effective vertices with
  two scales in the EFT for Multi-Regge processes in QCD}},  {\em Nucl. Phys.
  B} {\bf 946} (2019) 114715, [\href{http://xxx.lanl.gov/abs/1902.11030}{{\tt
  1902.11030}}].

\bibitem{Nefedov:2020ecb}
M.~A. Nefedov, {\it {Towards stability of NLO corrections in High-Energy
  Factorization via Modified Multi-Regge Kinematics approximation}},  {\em
  JHEP} {\bf 08} (2020) 055, [\href{http://xxx.lanl.gov/abs/2003.02194}{{\tt
  2003.02194}}].

\bibitem{Hentschinski:2020tbi}
M.~Hentschinski, K.~Kutak, and A.~van Hameren, {\it {Forward Higgs production
  within high energy factorization in the heavy quark limit at next-to-leading
  order accuracy}},  {\em Eur. Phys. J. C} {\bf 81} (2021), no.~2 112,
  [\href{http://xxx.lanl.gov/abs/2011.03193}{{\tt 2011.03193}}]. [Erratum:
  Eur.Phys.J.C 81, 262 (2021)].

\bibitem{Celiberto:2022fgx}
F.~G. Celiberto, M.~Fucilla, D.~Y. Ivanov, M.~M.~A. Mohammed, and A.~Papa, {\it
  {The next-to-leading order Higgs impact factor in the infinite top-mass
  limit}},  {\em JHEP} {\bf 08} (2022) 092,
  [\href{http://xxx.lanl.gov/abs/2205.02681}{{\tt 2205.02681}}].

\bibitem{Bergabo:2022zhe}
F.~Bergabo and J.~Jalilian-Marian, {\it {Single inclusive hadron production in
  DIS at small x: next to leading order corrections}},  {\em JHEP} {\bf 01}
  (2023) 095, [\href{http://xxx.lanl.gov/abs/2210.03208}{{\tt 2210.03208}}].

\bibitem{Taels:2023czt}
P.~Taels, {\it {Forward production of a Drell-Yan pair and a jet at small $x$
  at next-to-leading order}},  \href{http://xxx.lanl.gov/abs/2308.02449}{{\tt
  2308.02449}}.

\bibitem{Altinoluk:2023hfz}
T.~Altinoluk, N.~Armesto, A.~Kovner, and M.~Lublinsky, {\it {Single inclusive
  particle production at next-to-leading order in proton-nucleus collisions at
  forward rapidities: Hybrid approach meets TMD factorization}},  {\em Phys.
  Rev. D} {\bf 108} (2023), no.~7 074003,
  [\href{http://xxx.lanl.gov/abs/2307.14922}{{\tt 2307.14922}}].

\bibitem{vanHameren:2012if}
A.~van Hameren, P.~Kotko, and K.~Kutak, {\it {Helicity amplitudes for
  high-energy scattering}},  {\em JHEP} {\bf 1301} (2013) 078,
  [\href{http://xxx.lanl.gov/abs/1211.0961}{{\tt 1211.0961}}].

\bibitem{vanHameren:2016kkz}
A.~van Hameren, {\it {KaTie : For parton-level event generation with
  $k_T$-dependent initial states}},  {\em Comput. Phys. Commun.} {\bf 224}
  (2018) 371--380, [\href{http://xxx.lanl.gov/abs/1611.00680}{{\tt
  1611.00680}}].

\bibitem{Blanco:2020akb}
E.~Blanco, A.~van Hameren, P.~Kotko, and K.~Kutak, {\it {All-plus helicity
  off-shell gauge invariant multigluon amplitudes at one loop}},  {\em JHEP}
  {\bf 12} (2020) 158, [\href{http://xxx.lanl.gov/abs/2008.07916}{{\tt
  2008.07916}}].

\bibitem{Blanco:2022iai}
E.~Blanco, A.~Giachino, A.~van Hameren, and P.~Kotko, {\it {One-loop gauge
  invariant amplitudes with a space-like gluon}},  {\em Nucl. Phys. B} {\bf
  995} (2023) 116322, [\href{http://xxx.lanl.gov/abs/2212.03572}{{\tt
  2212.03572}}].

\bibitem{vanHameren:2022mtk}
A.~van Hameren, L.~Motyka, and G.~Ziarko, {\it {Hybrid k$_{T}$ -factorization
  and impact factors at NLO}},  {\em JHEP} {\bf 11} (2022) 103,
  [\href{http://xxx.lanl.gov/abs/2205.09585}{{\tt 2205.09585}}].

\bibitem{BermudezMartinez:2018fsv}
A.~Bermudez~Martinez, P.~Connor, H.~Jung, A.~Lelek, R.~\v{Z}leb\v{c}\'\i{}k,
  F.~Hautmann, and V.~Radescu, {\it {Collinear and TMD parton densities from
  fits to precision DIS measurements in the parton branching method}},  {\em
  Phys. Rev. D} {\bf 99} (2019), no.~7 074008,
  [\href{http://xxx.lanl.gov/abs/1804.11152}{{\tt 1804.11152}}].

\bibitem{Abdulov:2021ivr}
N.~A. Abdulov {\em et~al.}, {\it {TMDlib2 and TMDplotter: a platform for 3D
  hadron structure studies}},  {\em Eur. Phys. J. C} {\bf 81} (2021), no.~8
  752, [\href{http://xxx.lanl.gov/abs/2103.09741}{{\tt 2103.09741}}].

\bibitem{Catani:1998bh}
S.~Catani,
{\it {The Singular behavior of QCD amplitudes at two loop order}},
{\em Phys. Lett. B} {\bf 427} (1998), 161-171,
[\href{http://xxx.lanl.gov/abs/hep-ph/9802439}{{\tt hep-ph/9802439}}].

\bibitem{vanHameren:2010gg}
A.~van Hameren,
{\it {Kaleu: A General-Purpose Parton-Level Phase Space Generator}},
[\href{https://arxiv.org/abs/1003.4953}{{\tt arXiv:1003.4953}}].

\bibitem{Kleiss:1994qy}
R.~Kleiss and R.~Pittau,
{\it {Weight optimization in multichannel Monte Carlo}},
{\em Comput. Phys. Commun.} {\bf 83} (1994), 141-146,
[\href{https://arxiv.org/abs/hep-ph/9405257}{{\tt hep-ph/9405257}}].

\bibitem{Hou:2019qau}
T.~J.~Hou, K.~Xie, J.~Gao, S.~Dulat, M.~Guzzi, T.~J.~Hobbs, J.~Huston, P.~Nadolsky, J.~Pumplin and C.~Schmidt, \textit{et al.}
{\it {Progress in the CTEQ-TEA NNLO global QCD analysis}},
[\href{https://arxiv.org/abs/1908.11394}{{\tt arXiv:1908.11394}}].

\bibitem{Buckley:2014ana}
A.~Buckley, J.~Ferrando, S.~Lloyd, K.~Nordstr\"om, B.~Page, M.~R\"ufenacht, M.~Sch\"onherr and G.~Watt,
{\it {LHAPDF6: parton density access in the LHC precision era}},
{\em Eur. Phys. J. C} {\bf 75} (2015), 132,
[\href{https://arxiv.org/abs/1412.7420}{{\tt arXiv:1412.7420}}].

\bibitem{Kutak:2012rf}
K.~Kutak and S.~Sapeta,
{\it {Gluon saturation in dijet production in p-Pb collisions at Large Hadron Collider}},
{\em Phys. Rev. D} {\bf 86} (2012), 094043,
[\href{https://arxiv.org/abs/1205.5035}{{\tt arXiv:1205.5035}}].

\end{thebibliography}
\providecommand{\href}[2]{#2}\begingroup\raggedright\endgroup

\begin{appendix}
\addtocontents{toc}{\protect\setcounter{tocdepth}{1}}
\section{\label{App:splittingfunctions}Splitting functions}
Consider a collinear limit of momenta with arbitrary real momentum fractions $x,y$
%
\begin{equation}
p_a^\mu = xp^\mu + k_T^\mu - \frac{k_T^2}{2x\lop{p}{q}}\,q^\mu
\quad,\quad
p_b^\mu = yp^\mu - k_T^\mu - \frac{k_T^2}{2y\lop{p}{q}}\,q^\mu
\quad,\quad
k_T^\mu \to 0
\quad,
\end{equation}
%
where $p^\mu$ and $q^\mu$ are light-like with $\lop{p}{q}\neq0$, while $\lop{p}{k_T}=
\lop{q}{k_T}=0$.
The corresponding collinear splitting functions are defined as
%
\begin{align}
&\int\frac{d^{2-2\vepv}k_T}{\piep\mu^{-2\vepv}}\,\delta\big(\lambda-|k_T|^2\big)
\,\Mtree{}\big(p_a,p_b,\ldots\big)
\label{Eq:App15}\\&\hspace{24ex}
\;\overset{\lambda\to0}{\longrightarrow}\;
\frac{4\pi\alphaS}{\mu^{-2\vepv}}
\,\frac{\EuScript{P}\big(c(x+y)\to a(x)b(y)\big)}
       {\lop{p_a}{p_b}}
\,\Mtree{}\big((x+y)p,\ldots\big)
~\notag
\end{align}
%
and are given by
%
\begin{equation}
\EuScript{P}\big(c(x+y)\to a(x)b(y)\big)
=
\EuScript{Q}_{ab}(y/x)
=
c_{ab}\bar{\EuScript{Q}}_{ab}(y/x)
\end{equation}
%
with
%
\begin{align}
c_{gg} = 2\Nc \quad,\quad
\bar{\EuScript{Q}}_{gg}(\zeta) 
  &= \bigg[\frac{1}{\zeta}+\zeta+\frac{\zeta}{(1+\zeta)^2}\bigg]
~,\\
c_{qg} = 2C_F\quad,\quad
\bar{\EuScript{Q}}_{qg}(\zeta) 
  &= \bigg[\frac{1}{\zeta} + \frac{1-\vepv}{2}\frac{\zeta}{1+\zeta}\bigg]
~,\\
c_{gq} = 2C_F\quad,\quad
\bar{\EuScript{Q}}_{gq}(\zeta) 
  &= \sgn(\zeta)\bigg[\zeta + \frac{1-\vepv}{2}\frac{1}{1+\zeta}\bigg]
~,\\
c_{qq} = \frac{2T_R}{1-\vepv}\quad,\quad
\bar{\EuScript{Q}}_{qq}(\zeta) 
  &= \sgn(\zeta)\bigg[\frac{1-\vepv}{2}-\frac{\zeta}{(1+\zeta)^2}\bigg]
~.
\end{align}
%
Realize that both $\bar{\EuScript{Q}}_{ab}(\zeta)<0$ and $\lop{p_a}{p_b}<0$ for  $\zeta\in(-1,0)$, so the right-hand side of \Equation{Eq:App15} stays positive.
The functions satisfy
%
\begin{equation}
\bar{\EuScript{Q}}_{gg}(1/\zeta) = \bar{\EuScript{Q}}_{gg}(\zeta)
\;\;,\;\;
\bar{\EuScript{Q}}_{gq}(1/\zeta) = \sgn(\zeta)\bar{\EuScript{Q}}_{qg}(\zeta)
\;\;,\;\;
\bar{\EuScript{Q}}_{qq}(1/\zeta) = \bar{\EuScript{Q}}_{qq}(\zeta)
\;\;.
\end{equation}
%
Notice that for $z\in[0,1]$ we have
%
\begin{equation}
\bar{\EuScript{Q}}_{ab}\bigg(\frac{1-z}{z}\bigg) = \EuScript{P}_{ab}(z)/c_{ab}
\equiv \bar{\EuScript{P}}_{ab}(z)
~,
\end{equation}
%
where $\EuScript{P}_{ab}$ are the usual collinear splitting functions, but labelled following the splitting products rather than the parent, and such that there is a soft pole at $z=0$ if $a=g$, and at $z=1$ if $b=g$.
Notice also that
%
\begin{align}
\bar{\EuScript{Q}}_{gg}(\zeta) = -\frac{\bar{\EuScript{P}}_{gg}(1+\zeta)}{1+\zeta}
\quad&,\quad
\bar{\EuScript{Q}}_{qg}(\zeta) = -\frac{\bar{\EuScript{P}}_{qg}(1+\zeta)}{1+\zeta}
\quad,\\
\bar{\EuScript{Q}}_{gq}(\zeta) = \sgn(\zeta)\frac{\bar{\EuScript{P}}_{qq}(1+\zeta)}{1+\zeta}
\quad&,\quad
\bar{\EuScript{Q}}_{qq}(\zeta) = \sgn(\zeta)\frac{\bar{\EuScript{P}}_{gq}(1+\zeta)}{1+\zeta}
\quad.
\end{align}
%
Again for $z\in[0,1]$ this leads to 
%
\begin{align}
-z\EuScript{Q}_{ab}(z-1) = c_{ab}\bar{\EuScript{P}}_{\tau(ab)}(z)
\label{Eq:App84}\quad,
\end{align}
%
with
\begin{equation}
\tau(gg) = gg
\;\;,\;\;
\tau(qg) = qg
\;\;,\;\;
\tau(gq) = qq
\;\;,\;\;
\tau(qq) = gq
\;\;.\;\;
\end{equation}

\section{\label{App:correlators}Correlated matrix elements}
We present the correlated matrix elements that appear in the soft and collinear limits.
We present the color-correlated ones for completeness, and the spin-correlated ones in a form that is closer to what one would like to use in the context of helicity amplitudes.

Matrix elements involve a sum over color.
We can write the dependence of an amplitude on the color indices explicitly as
%
\begin{equation}
\Atree^{\omega_1\omega_2\cdots\omega_n}
~,
\end{equation}
%
where $\omega_l$ is the adjoint index $a_l$ if $l$ refers to a gluon, $\omega_l$ is the fundamental index $i_l$ if $l$ refers to a quark, and $\omega_l$ is the anti-fundamental index $j_l$ if $l$ refers to an anti-quark.
Implying the usual summation over repeated indices, the squared matrix element summed over color becomes
%
\begin{equation}
\big|\Atree\big|^2
=
\big(\Atree^{\bar{\omega}_1\bar{\omega}_2\cdots\bar{\omega}_n}\big)^*
\,\delta_{\bar{\omega}_1\omega_1}
  \delta_{\bar{\omega}_2\omega_2}
  \cdots
  \delta_{\bar{\omega}_n\omega_n}
\,\Atree^{\omega_1\omega_2\cdots\omega_n}
~.
\end{equation}
%
The color-correlated squared sum
%
\begin{equation}
\MtreeSumCor{k}{l}
\quad\textrm{is obtained by replacing}\quad
\delta_{\bar{\omega}_k\omega_k}
  \delta_{\bar{\omega}_l\omega_l}
\;\to\;
\EuScript{T}^{c}_{\bar{\omega}_k\omega_k}
\EuScript{T}^{c}_{\bar{\omega}_l\omega_l}
~,
\end{equation}
%
where
%
\begin{equation}
\EuScript{T}^{c}_{\bar{\omega}_l\omega_l}
=
\begin{cases}
\imag f^{\bar{a}_lca_l}\hspace{1ex}\textrm{if $l$ refers to a gluon,}\\
T^c_{\bar{\imath}_li_l}\hspace{3ex}\textrm{if $l$ refers to a quark,}\\
-T^c_{j_l\bar{\jmath}_l}\hspace{1.4ex}\textrm{if $l$ refers to an anti-quark.}
\end{cases}
\end{equation}
%
With this definition, the equal-index correlator $\MtreeSumCor{k}{k}$ is obtained, by only replacing $\delta_{\bar{\omega}_k\omega_k}$ with $\EuScript{T}^{c}_{\bar{\omega}_k\omega}\EuScript{T}^{c}_{\omega\omega_k}$.

For point-wise cancellation between collinear singularities in the real-radiation integral, spin correlations need to be included in the subtraction terms.
That is, we need the limit of \Equation{Eq:App15} before integration over the angular part of $k_T$, but only at $\vepv=0$.
The spin-correlated matrix elements necessary for the 4-dimensional subtraction terms can be written in terms of helicities as follows.
We can write the dependence of an amplitude on the helicity indices explicitly as
%
\begin{equation}
\Atree_{\lambda_1\lambda_2\cdots\lambda_n}
~.
\end{equation}
%
The squared amplitude summed over helicities becomes
%
\begin{equation}
\big|\Atree\big|^2 = \big(\Atree_{\bar{\lambda}_1\bar{\lambda}_2\cdots\bar{\lambda}_n}\big)^*
\,\delta^{\bar{\lambda}_1\lambda_1}
  \delta^{\bar{\lambda}_2\lambda_2}
  \cdots
  \delta^{\bar{\lambda}_n\lambda_n}
\,\Atree_{\lambda_1\lambda_2\cdots\lambda_n}
~.
\end{equation}
%
The $1$-helicity-fixed matrix element
%
\begin{equation}
\MtreeHelCor{l}
\quad\textrm{is obtained by replacing}\quad
\delta^{\bar{\lambda}_l\lambda_l}
\;\to\;
\delta^{\bar{\lambda}_l+}\delta^{\lambda_l-}
~.
\end{equation}
%
Let the helicity amplitude be constructed with polarization vectors $\varepsilon^{(l)}_{\lambda_l}$ for gluon $l$.
We denote
%
\begin{equation}
\EuScript{Q}_{lr}(\zeta)\otimes\big|\Atree\big|^2
=
\EuScript{Q}_{lr}(\zeta)\big|\Atree\big|^2
+c^{sc}_{lr}(\zeta)\,\frac{\zeta}{(1+\zeta)^2}\,\mathrm{Re}\Bigg\{\frac{\lop{\varepsilon^{(l)}_{+}}{p_r}}{\lop{\varepsilon^{(l)}_{-}}{p_r}}\,\MtreeHelCor{l}\Bigg\}
~,
\end{equation}
%
with 
%
\begin{equation}
c^{sc}_{gg}(\zeta)=4\Nc
\quad,\quad
c^{sc}_{qq}(\zeta)=-4T_R\,\sgn(\zeta)
\quad,\quad
c^{sc}_{gq}(\zeta)=c^{sc}_{qg}(\zeta)=0
\quad.
\end{equation}

\section{\label{App:limits}Integration limits}
For a function $f(x,y)$ that has support $0<x<1$ for the first argument, and vanishes outside this region we have
%
\begin{equation}
\int_0^1dx\int_{x-1}^xdy\,f(x-y,y)
=
\int_0^1dx\int_{-x}^{1-x}dy\,f(x,y)
~.
\end{equation}
%
To see this, firstly we observe that
%
\begin{align}
&\int_0^1dx\int_0^xdy\,f(x-y,y)\quad\graph{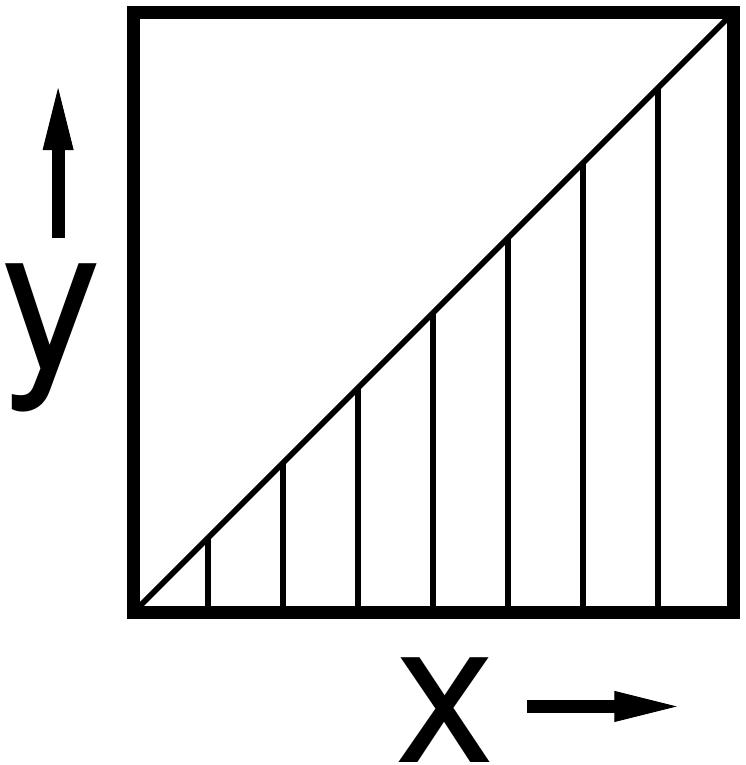}{6}{3}
\notag\\
= &\int_0^1dy\int_{y}^1dx\,f(x-y,y)\quad\graph{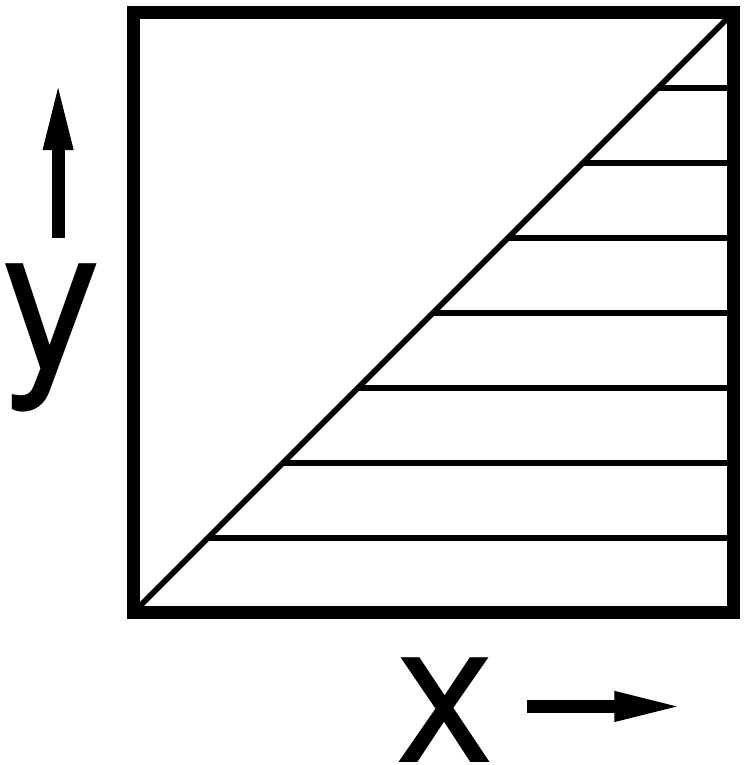}{6}{3}
\notag\\
= &\int_0^1dy\int_{0}^{1-y}dx\,f(x,y)\quad\graph{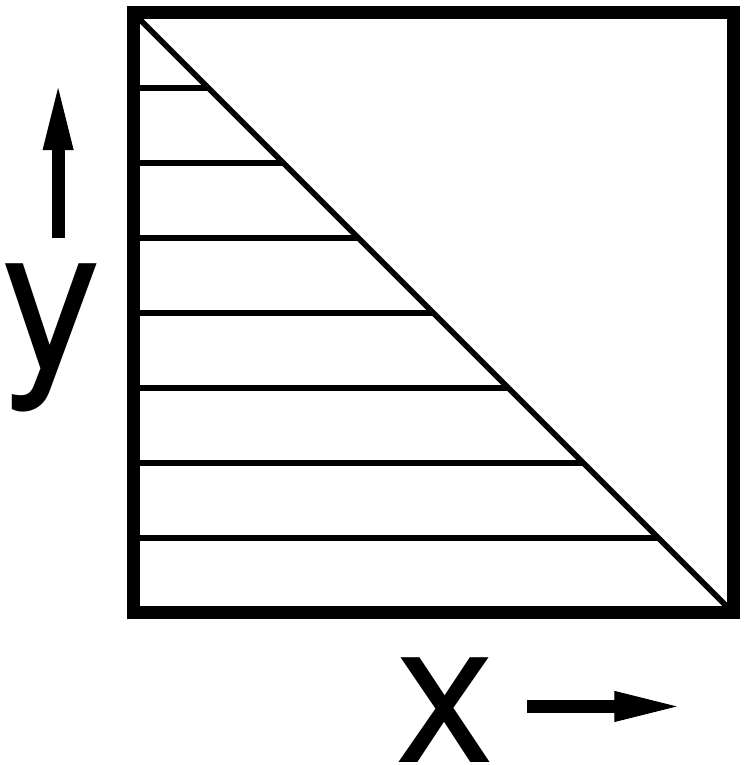}{6}{3}
\notag\\
= &\int_0^1dx\int_{0}^{1-x}dy\,f(x,y)\quad\graph{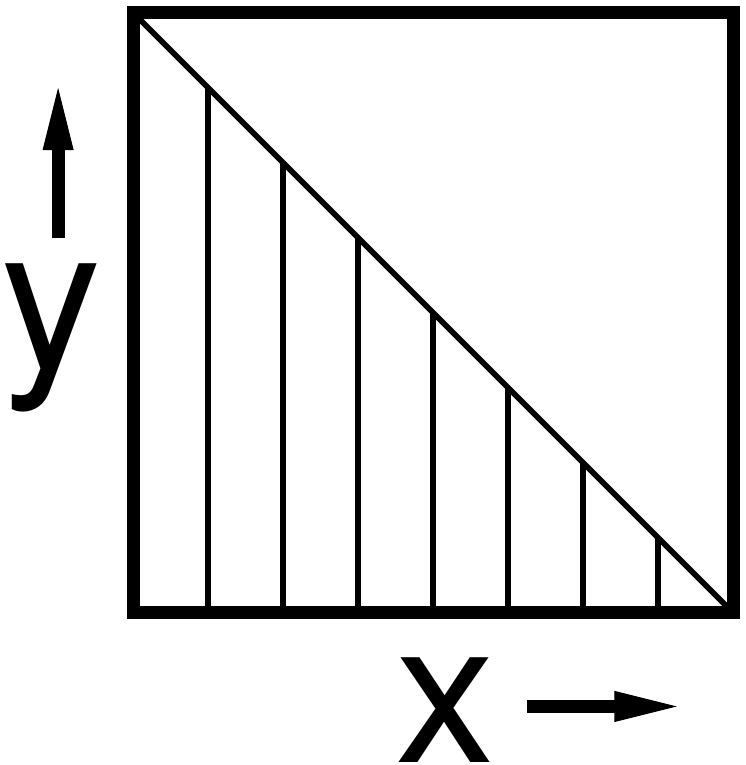}{6}{3}
~.
\end{align}
%
Going backwards, one finds
\begin{align}
\int_0^1dx\int_{x-1}^0dy\,f(x-y,y)
  & = \int_0^1dx\int_{0}^{1-x}dy\,f(x+y,-y)
\\
  &= \int_0^1dx\int_{0}^{x}dy\,f(x,-y)
   = \int_0^1dx\int_{-x}^{0}dy\,f(x,y)
~.
\end{align}

\section{\label{App:CollMap}Phase space mappings}
Here, we show that, for general function $h$,
%
\begin{align}
I
&\equiv
\int\dQin\int d\Phi\big(\Qin;\pSetNpls\big)
\,h\big(p_r\,,\,p_i\,,\,\Qin\,,\,\pSetNOri\big)
\label{Eq:App585}\\&=
  \int\dQin
  \int d\Phi\big(\Qin;\pSetNOr\big)
  \int\frac{d^4p_r}{(2\pi)^3}\delta_+(p_r^2)(1-z_{ri})
  \,h\big(p_r\,,\,(1-z_{ri})p_i\,,\,\Qin+p_r-z_{ri}p_i\,,\,\pSetNOr\big)
\notag~,
\end{align}
%
with $z_{ri}=E_r/E_i$ and with the implied integration limits
\begin{equation}
\theta\big(z_{ri}<1\big)
\,\theta\big({-}\xP<\xP_r-z_{ri}\xP_i<1-\xP\big)
\,\theta\big({-}\xM<\xM_r-z_{ri}\xM_i<1-\xM\big)
\end{equation}
%
on $p_r$.
First, we write
%
\begin{align}
I
&=
\int\dQin\int d\Phi\big(\Qin;\pSetNpls\big)
\,h\big(p_r\,,\,p_i\,,\,\Qin\,,\,\big\{(1+z_{ri})p_i,\pSetNOriB\big\}\big)
\notag~,
\end{align}
%
where we introduced the notation
%
\begin{equation}
\pSetNOri = \big\{(1+z_{ri})p_i,\pSetNOriB\big\}
\end{equation}
%
in order to single out the momentum $p_i$, and where $\pSetNOriB$ is $\pSetNpls$ with the momenta $p_r,p_i$ removed.
We can write the integration over the momenta $p_r$ and $p_i$ explicitly via
%
\begin{align}
I
&=
  \int\dQin
  \int\frac{d^4p_r}{(2\pi)^3}\delta_+(p_r^2)
  \int\frac{d^4p_i}{(2\pi)^3}\delta_+(p_i^2)
\\&\hspace{20ex}\times
  \int d\Phi\big(\Qin-p_r-p_i;\pSetNOriB\big)
  \,h\big(p_r\,,\,p_i\,,\,\Qin\,,\,\big\{(1+z_{ri})p_i,\pSetNOriB\big\}\big)
\notag~.
\end{align}
%
Now we change integration momentum $p_i$ to 
%
\begin{equation}
\tilde{p}_i=(1+z_{ri})p_i
\quad\Leftrightarrow\quad
p_i=(1-\tilde{z}_{ri})\tilde{p}_i
\quad\textrm{with}\quad
\tilde{z}_{ri}=\frac{E_r}{E_i+E_r}=\frac{z_{ri}}{1+z_{ri}}
~.
\end{equation}
%
Notice that
%
\begin{equation}
z_{ri}p_i = \tilde{z}_{ri}\tilde{p}_i
~.
\end{equation}
%
We get
%
\begin{align}
I
&=
  \int\dQin
  \int\frac{d^4p_r}{(2\pi)^3}\delta_+(p_r^2)
  \int\frac{d^4\tilde{p}_i}{(2\pi)^3}\delta_+(\tilde{p}_i^2)(1-\tilde{z}_{ri})
\\&\hspace{12ex}\times
  \int d\Phi\big(\Qin-p_r-(1-\tilde{z}_{ri})\tilde{p}_i;\pSetNOriB\big)
  \,h\big(p_r\,,\,(1-\tilde{z}_{ri})\tilde{p}_i\,,\,\Qin\,,\,\big\{\tilde{p}_i,\pSetNOriB\big\}\big)
\notag~.
\end{align}
%
The integration restriction $E_r<E_r+E_i=\tilde{E}_{i}\Leftrightarrow \tilde{z}_{ri}<1$ is from now on understood to be implied by the Jacobian $(1-\tilde{z}_{ri})$.
Now we rename $\tilde{p}_i\to p_i$
%
\begin{align}
I
&=
  \int\dQin
  \int\frac{d^4p_r}{(2\pi)^3}\delta_+(p_r^2)
  \int\frac{d^4p_i}{(2\pi)^3}\delta_+(p_i^2)(1-z_{ri})
\\&\hspace{12ex}\times
  \int d\Phi\big(\Qin-p_r+z_{ri}p_i-p_i;\pSetNOriB\big)
  h\big(p_r\,,\,(1-z_{ri})p_i\,,\,\Qin\,,\,\big\{p_i,\pSetNOriB\big\}\big)
\notag\\&=
  \int\dQin
  \int\frac{d^4p_r}{(2\pi)^3}\delta_+(p_r^2)
  \int d\Phi\big(\Qin-p_r+z_{ri}p_i;\pSetNOr\big)(1-z_{ri})
\\&\hspace{40ex}\times
  h\big(p_r\,,\,(1-z_{ri})p_i\,,\,\Qin\,,\,\pSetNOr\big)
\notag~.
\end{align}
%
Finally, we substitute $\Qin\leftarrow\Qin+p_r-z_{ri}p_i$ and use the result from \Appendix{App:limits} to find \Equation{Eq:App585}.
This was the ``collinear'' case.
The ``soft'' case is simpler, and we can see immediately that
%
\begin{align}
&\int\dQin\int d\Phi\big(\Qin;\pSetNpls\big)
\,h\big(p_r\,,\,\Qin\,,\,\pSetNOr\big)
\\&\hspace{16ex}=
  \int\dQin
  \int d\Phi\big(\Qin;\pSetNOr\big)
  \int\frac{d^4p_r}{(2\pi)^3}\delta_+(p_r^2)
  \,h\big(p_r\,,\,\Qin+p_r\,,\,\pSetNOr\big)
\notag~,
\end{align}
%
now with limits $\theta\big(0<\xP_r<1-\xP\big)\,\theta\big(0<\xM_r<1-\xM\big)$.

\section{Radiative integral representations}
We will encounter integrals of the type
%
\begin{equation}
I = \int d^{4}p_r\,\delta_+(p_r^2)\,f(p_r)
~.
\end{equation}
%
If $f$ depends on a an external momentum $p_i$, it is useful to use a frame with $p_i$ along the $z$-axis:
%
\begin{equation}
I = 
  \int_0^\infty\hspace{-1ex}dE_rE_r
  \int_{0}^1\hspace{0ex}d\zeta
  \int_0^{2\pi}\hspace{-1ex}d\varphi
  \,f(p_r;p_i)
\end{equation}
%
with
\begin{equation}
p_{r0} = E_r
\;\;,\quad
\vec{p}_r = 2E_r\,\mathrm{Rot}_i\Big(\sqrt{\zeta-\zeta^2}\cos(\varphi),\sqrt{\zeta-\zeta^2}\sin(\varphi),\srac{1}{2}-\zeta\Big)
~,
\end{equation}
%
and where 
\begin{equation}
\textrm{%
$\mathrm{Rot}_i$ is the inverse of the rotation that rotates $\vec{n}_i$ to the $z$-axis: $\mathrm{Rot}_i(0,0,1)=\vec{n}_i$.
}
\end{equation}
%
We also encounter the $(4+\vep)$-dimensional case, but with $f$ only depending on $p_i$ via $\lop{p_i}{p_r}$:
%
%
\begin{equation}
\int d^{4+\vep}p_r\,\delta_+(p_r^2)\,f\big(E_r,\lop{p_i}{p_r}\big) = 
  2^{\vep+1}\piep
  \int_0^\infty\hspace{-1ex}dE_rE_r^{1+\vep}
  \int_{0}^1\hspace{0ex}d\zeta\,\zeta^{\vep/2}(1-\zeta)^{\vep/2}
  \,f\big(E_r,2E_rE_i\zeta\big)
~.
\end{equation}
%

When there are two inner products $\lop{p_i}{p_r},\lop{p_j}{p_r}$ involved, it can be useful to apply the Sudakov decomposition
%
%
\begin{equation}
p_r^\mu = y_ip_i^\mu + y_jp_j^\mu + y_T^\mu
~,
\end{equation}
%
with
%
\begin{equation}
y_i=\frac{2\lop{p_j}{p_r}}{s_{ij}}
\;\;,\;\;              
y_j=\frac{2\lop{p_i}{p_r}}{s_{ij}}
\;\;,\;\;
\lop{p_i}{y_T}=\lop{p_j}{y_T}=0
\;\;,\;\;
s_{ij}=2\lop{p_i}{p_j}
~,
\end{equation}
%
so
%
\begin{equation}
\int d^{4+\vep}p_r\,\delta_+(p_r^2)\,f\big(p_r\big) = 
\frac{s_{ij}}{2}
\int_0^\infty\hspace{-1ex}dy_i
\int_0^\infty\hspace{-1ex}dy_j
\int d^{2+\vep}y_T\,\delta_+\big(y_iy_js_{ij}-|y_T|^2\big)
\,f(p_r)
~.
\end{equation}
%
If the integrand does not depend on the direction of $y_T$, we can evaluate further to
%
%
\begin{align}
\int d^{4+\vep}p_r\,\delta_+(p_r^2)
\,f\big(y_i,y_j,|y_T|\big)
&=
\frac{\piep\,s_{ij}^{1+\vep/2}}{2}
\int_0^\infty\hspace{-1ex}dy_i\,y_i^{\vep/2}
\int_0^\infty\hspace{-1ex}dy_j\,y_j^{\vep/2}
\,f\big(y_i,y_j,\sqrt{y_iy_js_{ij}}\big)
~.
\end{align}
%

\section{\label{App:divergent}Calculation of the divergent integrals}
We encounter the following basic integrals
%
\begin{align}
\int_0^\infty\hspace{-1ex}dx\,x^{\vep/2-1}(1+x)^{-1-\vep}
&=
\frac{2}{\vep}-\vep\,\frac{\pi^2}{12}+\Ord\big(\vep^2\big)
\\
\int_0^1\hspace{-0ex}dx\,x^{\vep/2-1}(1+zx)^{-\vep}
&= \frac{2}{\vep} + \vep\mathrm{Li}_2(-z) + \Ord\big(\vep^2\big)
\\
\int_{0}^{z}dx\,x^{\vep/2-1}\,(1-x)^{\vep/2}
&=
\frac{2}{\vep}
+\ln(z)+\vep\bigg(\frac{1}{4}\ln^2(z)-\frac{1}{2}\mathrm{Li}_2(z)\bigg)+\Ord\big(\vep^2\big)
\\
\int_0^{1}\hspace{0ex}dx\,x^{\vep-1}(1-zx)
&
 = \frac{1}{\vep} -z + \vep z
  +\Ord\big(\vep^2\big)
\\
\int_0^z dx\,x^{\vep-1}
&
  = \frac{1}{\vep} + \ln(z) + \frac{\vep}{2}\ln^2(z)
  +\Ord\big(\vep^2\big)
\\
\int_0^z dx\,x^{\vep+n}
&
  = \frac{z^{n+1}}{n+1}\bigg\{1+\vep\bigg[\ln(z)-\frac{1}{n+1}\bigg]\bigg\}
  +\Ord\big(\vep^2\big)
\\
\int_0^z dx\,\frac{x^{\vep}}{1-x}
&=
  -\ln(1-z)
  +\vep\bigg[\mathrm{Li}_2(1-z)-\frac{\pi^2}{6}\bigg]
  +\Ord\big(\vep^2\big)
\\
\int_{0}^{1}dx\,x^{\vep/2-1}\,(1-x)^{\vep/2}
&=
\frac{2}{\vep} - \vep\,\frac{\pi^2}{12} + \Ord\big(\vep^2\big)
\\
\int_{0}^{1}dx\,x^{\vep/2-1}\,(1-x)^{\vep/2-1}
&=
\frac{4}{\vep} - \vep\,\frac{\pi^2}{6} + \Ord\big(\vep^2\big)
\\
\int_0^{\infty}dx\,x^{\vep/2-1}(1+x)^{-\vep}
&=
2\int_0^{1}dx\,x^{\vep/2-1}(1+x)^{-\vep}
 = \frac{4}{\vep} - \vep\,\frac{\pi^2}{6} + \Ord\big(\vep^2\big)
\end{align}
%

\subsection{Final-state}
\subsubsection{Collinear}
\begin{align}
\LXYZ{\Fcoldiv}{i}
&=
\frac{1}{\piep\mu^{\vep}}
\int d^{4+\vep}p_r\,\delta_+(p_r^2)
\,\frac{\theta(\lop{n_r}{n_i}<2\zeta_0)\,\theta(z_{ri}<\srac{1}{2})}
       {\lop{p_i}{p_r}}
\,\EuScript{P}_{ri}(1-z_{ri})
%
\notag\\&\hspace{0ex}=
\frac{2^{\vep+1}}{\mu^{\vep}}
\,\int_0^\infty dE_r\,E_r^{1+\vep}
\int_{0}^{\zeta_0} d\zeta\,\zeta^{\vep/2}(1-\zeta)^{\vep/2}
\,\frac{\theta\big(z_{ri}<\srac{1}{2}\big)}{2E_rE_i\zeta}
\,\EuScript{P}_{ir}(1-z_{ri})
\notag\\&\hspace{0ex}=
\bigg(\frac{2E_i}{\mu}\bigg)^{\!\vep}
\int_0^{1/2}dx\,x^{\vep}
\,\EuScript{P}_{ir}(1-x)
\int_{0}^{\zeta_0} d\zeta\,\zeta^{\vep/2-1}(1-\zeta)^{\vep/2}
\end{align}

\subsubsection{Soft-collinear}
\begin{align}
\frac{\LXYZ{\Fsoftcoldiv}{i}}{-2C_i}
&=
\,\frac{1}{\piep\mu^\vep}
\int d^{4+\vep}p_r\,\delta_+(p_r^2)
\,\theta(E_r<E_0)\,\theta(\lop{n_i}{n_r}<2\zeta_0)(1-z_{ri})
\,\frac{1}{E_r^2\,\lop{n_i}{n_r}}
\notag\\&\hspace{0ex}=
\frac{2^{1+\vep}}{\mu^\vep}
\int_0^{E_0}\hspace{-1ex}dE_rE_r^{\vep+1}
\int_{0}^{\zeta_0}d\zeta\,\zeta^{\vep/2}(1-\zeta)^{\vep/2}
\,(1-z_{ri})
\,\frac{1}{2E_r^2\zeta}
\notag\\&\hspace{0ex}=
\frac{2^\vep}{\mu^\vep}
\int_0^{E_0}\hspace{-1ex}dE_r\,E_r^{\vep-1}(1-E_r/E_i)
\int_{0}^{\zeta_0}d\zeta\,\zeta^{\vep/2-1}\,(1-\zeta)^{\vep/2}
\notag\\&\hspace{0ex}=
\bigg(\frac{2E_0}{\mu}\bigg)^{\!\vep}
\int_0^{1}d\eta\,\eta^{\vep-1}(1-z_{0i}\eta)
\int_{0}^{\zeta_0}d\zeta\,\zeta^{\vep/2-1}\,(1-\zeta)^{\vep/2}
~.
\end{align}

\subsubsection{Soft}
%
Also here it is understood that $E_0$ is in fact the minimum of $E_0$ and $E_i$.
In the following, $i$ refers to a final-state momentum, while $j$ can both be initial-state and final-state.
%
\begin{align}
&\frac{\LXYZ{\Fsoftdiv}{ij}}{-2}
=
\frac{1}{\piep\mu^\vep}\int d^{4+\vep}p_r\,\delta_+(p_r^2)
\,\theta\big(E_r^{(ij)}<E_0\big)\,\bigg(1-\frac{E_r^{(ij)}}{E_i}\bigg)
\,\frac{1}{\lop{n_i}{p_r}}\,\frac{\lop{n_i}{n_j}}{\lop{n_i}{p_r}+\lop{n_j}{p_r}}
\notag\\&\hspace{0ex}=
\,\frac{1}{\piep\mu^\vep}\int d^{4+\vep}p_r\,\delta_+(p_r^2)
\,\theta\big(E_r^{(ij)}<E_0\big)\,\bigg(1-\frac{E_r^{(ij)}}{E_i}\bigg)
\,\frac{1}{\lop{p_i}{p_r}}
\,\frac{\lop{p_i}{p_j}}{z_{ji}\lop{p_r}{p_i}+\lop{p_r}{p_j}}
\quad\bigg[y_{j/i}=\frac{\lop{p_r}{p_{i/j}}}{\lop{p_i}{p_j}}\bigg]
\notag\\&\hspace{0ex}=
\bigg(\frac{s_{ij}}{\mu^2}\bigg)^{\!\vep/2}
\int_0^\infty\hspace{-1ex}dy_i\,y_i^{\vep/2}
\int_0^\infty\hspace{-1ex}dy_j\,y_j^{\vep/2}
\,\theta\big(y_iE_i+y_jE_j<E_0\big)
\,\frac{1}{y_j}
 \,\frac{1-y_i-z_{ji}y_j}{y_i+z_{ji}y_j}
\quad\big[y_j=y_i\zeta\big]
\notag\\&\hspace{0ex}=
\bigg(\frac{s_{ij}}{\mu^2}\bigg)^{\!\vep/2}
\int_0^\infty\hspace{-1ex}dy_i\,y_i^{\vep-1}
\int_0^\infty\hspace{-1ex}d\zeta\,\zeta^{\vep/2-1}
\,\theta\big(y_i(1+z_{ji}\zeta)<z_{0i}\big)
 \,\frac{1-y_i(1+z_{ji}\zeta)}{1+z_{ji}\zeta}
\quad\bigg[y_i =\frac{\eta}{1+z_{ji}\zeta}\bigg]
\notag\\&\hspace{0ex}=
\bigg(\frac{s_{ij}}{\mu^2}\bigg)^{\!\vep/2}
\int_0^{z_{0i}}\hspace{-1ex}d\eta\,\eta^{\vep-1}(1-\eta)
\int_0^\infty\hspace{-1ex}d\zeta\,\zeta^{\vep/2-1}
\,(1+z_{ji}\zeta)^{-\vep}
\,\frac{1}{1+z_{ji}\zeta}
\quad\big[\eta\leftarrow z_{0i}\eta\;,\;\;\zeta\leftarrow z_{ij}\zeta\big]
\notag\\&\hspace{0ex}=
\bigg(\frac{s_{ij}E_0^2}{\mu^2E_iE_j}\bigg)^{\!\vep/2}
\int_0^{1}\hspace{0ex}d\eta\,\eta^{\vep-1}(1-z_{0i}\eta)
\int_0^\infty\hspace{-1ex}d\zeta\,\zeta^{\vep/2-1}\,(1+\zeta)^{-1-\vep}
%
%
%
\end{align}
%

\subsection{Initial-state terms}

In the following, $i$ refers to an initial-state momentum, while $j$ can both be initial-state and final-state.
\begin{align}
\frac{\LXYZ{\Isoftdiv}{ij}}{-2}
&=
\,\frac{1}{\piep\mu^\vep}\int d^{4+\vep}p_r\,\delta_+(p_r^2)
\,\theta\big(E_r^{(ij)}<E_0\big)\,\frac{1}{\lop{n_i}{p_r}}
\,\frac{\lop{n_i}{n_j}}{\lop{p_r}{n_i}+\lop{p_r}{n_j}}
\notag\\&\hspace{0ex}=
\bigg(\frac{s_{ij}E_0^2}{\mu^2E_iE_j}\bigg)^{\!\vep/2}
\int_0^{1}\hspace{0ex}d\omega\,\omega^{\vep-1}
\int_0^\infty\hspace{-1ex}d\zeta\,\zeta^{\vep/2-1}\,(1+\zeta)^{-1-\vep}
\end{align}
%
%
\begin{align}
\frac{\LXYZ{\Isoftcoldiv}{\inlbl}}{-2C_{\inlbl}}
&=
\frac{1}{\piep\mu^\vep}
\,\frac{2}{\sTot}
\int d^{4+\vep}p_r\,\delta_+(p_r^2)
\,\theta(E_r<E_0)\,\theta\big(\xM_r<\xiP\xP_r\big)
\,\frac{1}{\xP_r\xM_r}
\notag\\&\hspace{0ex}=
\bigg(\frac{\sTot}{\mu^2}\bigg)^{\!\vep/2}
\int_0^{\infty}d\xP_r\xP_r^{\vep/2-1}\int_0^{\infty}d\xM_r\xM_r^{\vep/2-1}
\,\theta\big(E\xP_r+\bar{E}\xM_r<E_0\big)\,\theta\big(\xM_r<\xiP\xP_r\big)
\notag\\&\hspace{0ex}=
\bigg(\frac{\sTot}{\mu^2}\bigg)^{\!\vep/2}
\int_0^{\infty}d\xP_r\xP_r^{\vep-1}\int_0^{\infty}d\zM\,\zM^{\vep/2-1}
\,\theta\big(\xP_r(E+\bar{E}\zM)<E_0\big)\,\theta\big(\zM<\xiP\big)
\notag\\&\hspace{0ex}=
\bigg(\frac{\sTot}{\mu^2}\bigg)^{\!\vep/2}
\int_0^{\infty}d\zP\,\zP^{\vep-1}\int_0^{\infty}d\zM\,\zM^{\vep/2-1}
(E+\bar{E}\zM)^{-\vep}
\,\theta\big(\zP<E_0\big)\,\theta\big(\zM<\xiP\big)
\notag\\&\hspace{0ex}=
\bigg(\frac{\sTot E_0^2\xiP}{\mu^2E^2}\bigg)^{\!\vep/2}
\int_0^{1}d\omega\,\omega^{\vep-1}\int_0^{1}d\zeta\,\zeta^{\vep/2-1}
\bigg(1+\frac{\bar{E}\xiP}{E}\zeta\bigg)^{-\vep}
~,
\end{align}

\begin{align}
\frac{\LXYZ{\Isoftdiv,2}{\inlbl}}{-2}
&=
\,\frac{1}{\piep\mu^\vep}\int d^{4+\vep}p_r\,\delta_+(p_r^2)
\,\theta(E_r<E_0)\,\frac{1}{E_r^2\,\lop{n_{\inlbl}}{n_r}}
\,\theta(1-\xP<E_r/E)
\notag\\&\hspace{0ex}=
\frac{2^{1+\vep}}{\mu^\vep}
\int_0^{\infty}\hspace{-1ex}dE_rE_r^{\vep+1}
\int_{0}^{1}d\zeta\,\zeta^{\vep/2}(1-\zeta)^{\vep/2}
\,\frac{1}{2E_r^2\zeta}
\,\theta\bigg(\frac{1-\xP}{E_0/E}<E_r/E_0<1\bigg)
\notag\\&\hspace{0ex}=
\bigg(\frac{2E_0}{\mu}\bigg)^{\!\vep}
\int_0^{1}\hspace{0ex}d\omega\,\omega^{\vep-1}
\,\theta\bigg(\frac{1-\xP}{E_0/E}<\omega\bigg)
\int_{0}^{1}d\zeta\,\zeta^{\vep/2-1}\,(1-\zeta)^{\vep/2}
\end{align}
%
%
\begin{align}
\frac{\LXYZ{\Isoftcoldiv,2}{\inlbl}}{-2C_{\inlbl}}
&=
\frac{1}{\piep\mu^\vep}
\,\frac{2}{\sTot}
\int d^{4+\vep}p_r\,\delta_+(p_r^2)
\,\theta(E_r<E_0)\,\theta\big(\xM_r<\xiP\xP_r\big)
\,\theta(1-\xP<E_r/E)
\,\frac{1}{\xP_r\xM_r}
\notag\\&\hspace{0ex}=
\bigg(\frac{\sTot}{\mu^2}\bigg)^{\!\vep/2}
\int_0^{\infty}\hspace{-1ex}d\xP_r\xP_r^{\vep/2-1}
\int_0^{\infty}\hspace{-1ex}d\xM_r\xM_r^{\vep/2-1}
\,\theta\big(\xM_r<\xiP\xP_r\big)
\,\theta\big(1-\xP<\xP_r+\xM_r\bar{E}/E<E_0/E\big)
\notag\\&\hspace{0ex}=
\bigg(\frac{\sTot}{\mu^2}\bigg)^{\!\vep/2}
\int_0^{\infty}\hspace{-1ex}d\xP_r\xP_r^{\vep-1}
\int_0^{\infty}\hspace{-1ex}d\zM\,\zM^{\vep/2-1}
\,\theta\big(\zM<\xiP\big)
\,\theta\big(1-\xP<\xP_r(1+\zM\bar{E}/E)<E_0/E\big)
\notag\\&\hspace{0ex}=
\bigg(\frac{\sTot}{\mu^2}\bigg)^{\!\vep/2}
\int_0^{\infty}\hspace{-1ex}d\zP\,\zP^{\vep-1}
\int_0^{\xiP}d\zM\,\zM^{\vep/2-1}
(1+\zM\bar{E}/E)^{-\vep}
\,\theta\big(1-\xP<\zP<E_0/E\big)
\notag\\&\hspace{0ex}=
\bigg(\frac{\sTot E_0^2\xiP}{\mu^2E^2}\bigg)^{\!\vep/2}
\int_0^{1}d\omega\,\omega^{\vep-1}
\,\theta\bigg(\frac{1-\xP}{E_0/E}<\omega\bigg)
\int_0^{1}d\zeta\,\zeta^{\vep/2-1}\bigg(1+\frac{\bar{E}\xiP}{E}\,\zeta\bigg)^{\!-\vep}
\end{align}

\begin{align}
\frac{\LXYZ{\Isoftcoldiv,2}{\inlbl}}{-2C_{\inlbl}}
-\frac{\LXYZ{\Isoftdiv,2}{\inlbl}}{-2}
&=
-\ln\bigg(\frac{1-x}{E_0/E}\bigg)\ln\bigg(\frac{\bar{E}\xiP}{E}\bigg)
 \theta\bigg(\frac{1-\xP}{E_0/E}<1\bigg)
+\Ord(\vep)
\end{align}

%
%

\section{\label{App:finite}Massaging the finite integrals}
We bring the integrals to a form
%
\begin{equation}
\int_0^1d\rho_1\int_0^1d\rho_2\int_0^1d\rho_3\,f(\rho_1,\rho_2,\rho_3)
\,\frac{g(\rho_1,\rho_2,\rho_3)-1}{\rho_1\rho_2}
~,
\end{equation}
%
where $f,g$ are finite and $\rho_1=0,\rho_2=0$ represent the soft and collinear singularities.
One frequently applied ingredient will be the variable substitution
%
\begin{equation}
\int_0^1d\eta\,f(\rho_1,\eta,\rho_3)
\,\frac{g(\rho_1,\eta,\rho_3)-1}{\rho_1\eta}
=
\int_0^1d\rho_2\,2f(\rho_1,\rho_2^2,\rho_3)
\,\frac{g(\rho_1,\rho_2^2,\rho_3)-1}{\rho_1\rho_2}
\end{equation}
%
when the function $g$ found at some stage turns out to behave as
\begin{equation}
g(\rho_1,\eta,\rho_3)=1+\rho_1\sqrt{\eta}\,g_{1,1}(\rho_3)+\cdots
~.
\end{equation}

%
\subsection{Final-state terms}
\subsubsection{Collinear}
We need to calculate
\begin{align}
\LXYZ{\Fcolfin}{i}
&=
\frac{1}{\pi}
\int d^{4}p_r\,\delta_+(p_r^2)
\,\SXYZ{\Fcol}{}(p_r;p_i)
\,\Big[\ELshiftPerp{p_{r}-z_{ri}p_{i}}\,\Theta(p_{r}-z_{ri}p_{i})-1\Big]
\end{align}
with
\begin{equation}
\SXYZ{\Fcol}{}(p_r;p_i)
=
\tilde{\EuScript{P}}_{ir}(z_{ri})
\,\frac{\theta(\lop{n_r}{n_i}<2\zeta_0)\,\theta(z_{ri}<\srac{1}{2})}
       {E_r^2\,\lop{n_i}{n_r}}
~,
\end{equation}
%
and where
%
\begin{equation}
\tilde{\EuScript{P}}_{ir}(z) = 
z\,\EuScript{P}_{ir}(1-z)
\end{equation}
%
is finite at $z=0$.
Simply using a frame in which $p_i$ is along the $z$-axis, we can write
%
\begin{align}
\LXYZ{\Fcolfin}{i}
&=
\frac{1}{\pi}
\,\int_0^{1/2}\hspace{0ex}\frac{d\omega}{\omega}
\,\tilde{\EuScript{P}}_{ir}(\omega)
\int_{0}^{\zeta_0}\hspace{0ex}\frac{d\zeta}{2\zeta}
\int_0^{2\pi} d\varphi
\,\Big[\ELshiftPerp{\rc}\,\Theta(\rc)-1\Big]
~,
\end{align}
with
\begin{equation}
\rc_0 = 0
\quad,\quad
\vec{\rc} = 2E_i\omega\,\mathrm{Rot}_i\Big(\sqrt{\zeta-\zeta^2}\cos(\varphi),\sqrt{\zeta-\zeta^2}\sin(\varphi),-\zeta\Big)
~.
\end{equation}
%
%
Substituting $\zeta=\eta^2$:
%
\begin{align}
\LXYZ{\Fcolfin}{i}
&=
\frac{1}{\pi}
\,\int_0^{1/2}\hspace{0ex}\frac{d\omega}{\omega}
\,\tilde{\EuScript{P}}_{ir}(\omega)
\int_{0}^{\sqrt{\zeta_0}}\hspace{0ex}\frac{d\eta}{\eta}
\int_0^{2\pi} d\varphi
\,\Big[\ELshiftPerp{\rc}\,\Theta(\rc)-1\Big]
~,
\end{align}
with
\begin{equation}
\vec{\rc} = 2E_i\omega \eta\,\mathrm{Rot}_i\Big(\sqrt{1-\eta^2}\cos(\varphi),\sqrt{1-\eta^2}\sin(\varphi),-\eta\Big)
~.
\end{equation}
%
Normalizing the variables
%
\begin{align}
\LXYZ{\Fcolfin}{i}
&=
2
\,\int_{0}^{1}\hspace{0ex}d\rho_1
\int_{0}^{1}\hspace{0ex}d\rho_2
\int_{0}^{1}\hspace{0ex}d\rho_3
\,\tilde{\EuScript{P}}_{ir}\big(\rho_1/2\big)
\,\frac{\ELshiftPerp{\rc}\,\Theta(\rc)-1}{\rho_1\rho_2}
~,
\end{align}
with
\begin{equation}
\vec{\rc} = E_i\sqrt{\zeta_0}\,\rho_1\rho_2\,\mathrm{Rot}_i\Big(\sqrt{1-\zeta_0\rho_2^2}\cos(2\pi\rho_3),\sqrt{1-\zeta_0\rho_2^2}\sin(2\pi\rho_3),-\sqrt{\zeta_0}\,\rho_2\Big)
~.
\end{equation}
One could consider substituting
\begin{equation}
\rho_1 = e^{(1-\tau_2)\ln(\tau_1)}
\;\;,\;\;
\rho_2 = e^{\tau_2\ln(\tau_1)}
\quad\Longleftrightarrow\quad
\tau_1=\rho_1\rho_2
\;\;,\;\;
\tau_2=\frac{\ln(\rho_1)}{\ln(\rho_1\rho_2)}
~,
\end{equation}
%
so
%
\begin{align}
\LXYZ{\Fcolfin}{i}
&=
2
\,\int_{0}^{1}\hspace{0ex}d\tau_1
\int_{0}^{1}\hspace{0ex}d\tau_2
\int_{0}^{1}\hspace{0ex}d\tau_3
\,\tilde{\EuScript{P}}_{ir}\big(\rho_1/2\big)
\ln\bigg(\frac{1}{\tau_1}\bigg)
\frac{\ELshiftPerp{\rc}\,\Theta(\rc)-1}{\tau_1}
~,
\end{align}
and see explicitly that there is only one singular variable, and one subtraction is enough to regularize the integral.

\subsubsection{Soft}
%
We need to calculate
%
\begin{align}
\LXYZ{\Fsoftfin}{ib}
&=
\frac{-2}{\pi}
\int d^{4}p_r\,\delta_+(p_r^2)
\,\frac{1}{E_r^2\,\lop{n_i}{n_r}}
\,A_r^{(ib)}
\,F\big(p_{r}-z_{ri}p_{i},E_r,A_r^{(ib)}\big)
\end{align}
%
with
%
\begin{align}
A_r^{ib} = \frac{\lop{n_i}{n_b}}{\lop{n_i}{n_r}+\lop{n_b}{n_r}}
~,
\end{align}
%
and
\begin{align}
F\big(p_{r}-z_{ri}p_{i},E_r,A_r^{(ib)}\big)
&=
\ELshiftPerp{p_r-z_{ri}p_i}\,\Theta(p_r-z_{ri}p_i)
\,\theta\big(E_r<E_0\big)\bigg(1-\frac{E_r}{E_i}\bigg)
\notag\\&\hspace{24ex}
-\theta\big(E_r<E_0A_r^{(ib)}\big)\bigg(1-\frac{E_r}{E_iA_r^{(ib)}}\bigg)
~.
\end{align}
%
The maximum value of $A_r^{(ib)}$ is $2$, which can be reached for $n_r\sim n_i+n_b$ and $\lop{n_i}{n_b}\to0$, so using a frame in which $p_i$ is along the $z$-axis, we can write
%
\begin{align}
\LXYZ{\Fsoftfin}{i}
&=
\frac{-2}{\pi}
\,\int_0^{2E_0}\hspace{0ex}\frac{dE_r}{E_r}
\int_{0}^{1}\hspace{0ex}\frac{d\zeta}{2\zeta}
\int_0^{2\pi} d\varphi
\,A_r^{(ib)}
\,F\big(v,E_r,A_r^{(ib)}\big)
~,
\end{align}
with
\begin{align}
\rc_0=0\quad,\quad
\vec{\rc} &= 2E_r\,\mathrm{Rot}_i\Big(\sqrt{\zeta-\zeta^2}\cos(\varphi),\sqrt{\zeta-\zeta^2}\sin(\varphi),-\zeta\Big)
\quad,
\\
\vec{n}_{r} &= \vec{\rc}/E_r + \vec{n}_i
\quad,\quad \lop{n_i}{n_r} = 2\zeta
~.
\end{align}
%
Substituting $\zeta=\rho_2^2$ and normalizing the other variables, we get
%
\begin{align}
\LXYZ{\Fsoftfin}{i}
&=
-4
\,\int_{0}^{1}\hspace{0ex}d\rho_1
\int_{0}^{1}\hspace{0ex}d\rho_2
\int_{0}^{1}\hspace{0ex}d\rho_3
\,A_r^{(ib)}\,\frac{F\big(v,2E_0\rho_1,A_r^{(ib)}\big)}{\rho_1\rho_2}
~,
\end{align}
with
\begin{align}
v_0=0\quad,\quad
\vec{\rc} &= 4E_0\rho_1\rho_2\,\mathrm{Rot}_i\Big(\sqrt{1-\rho_2^2}\cos(2\pi\rho_3),\sqrt{1-\rho_2^2}\sin(2\pi\rho_3),-\rho_2\Big)
\quad,
\\
\vec{n}_{r} &= \vec{\rc}/(2E_0\rho_1) + \vec{n}_i
\quad,\quad \lop{n_i}{n_r} = 2\rho_2^2
~.
\end{align}

\subsubsection{Soft-collinear}
%
We need to calculate
%
\begin{align}
\LXYZ{\Fsoftcolfin}{i}
&=
\frac{-2C_i}{\pi}
\int d^{4}p_r\,\delta_+(p_r^2)
\,\frac{\theta(E_r<E_0)\theta(\lop{n_i}{n_r}<2\zeta_0)}{E_r^2\,\lop{n_i}{n_r}}
\,F(p_{r}-z_{ri}p_{i},E_r)
\end{align}
%
with
%
\begin{align}
F\big(p_{r}-z_{ri}p_{i},E_r\big)
&=
\bigg(1-\frac{E_r}{E_i}\bigg)
\Big[
\ELshiftPerp{p_r-z_{ri}p_i}\,\Theta(p_r-z_{ri}p_i) - 1\Big]
~.
\end{align}
Simply using a frame in which $p_i$ is along the $z$-axis, we can write
%
\begin{align}
\LXYZ{\Fsoftfin}{i}
&=
\frac{-2C_i}{\pi}
\,\int_0^{E_0}\hspace{0ex}\frac{dE_r}{E_r}
\int_{0}^{\zeta_0}\hspace{0ex}\frac{d\zeta}{2\zeta}
\int_0^{2\pi} d\varphi
\,F(v,E_r)
~,
\end{align}
with
\begin{align}
\rc_0=0\quad,\quad
\vec{\rc} &= 2E_r\,\mathrm{Rot}_i\Big(\sqrt{\zeta-\zeta^2}\cos(\varphi),\sqrt{\zeta-\zeta^2}\sin(\varphi),-\zeta\Big)
~.
\end{align}
%
Substituting $\zeta=\eta^2$ and normalizing the variables, we get
%
%
\begin{align}
\LXYZ{\Fsoftcolfin}{i}
&=
-4C_i
\,\int_{0}^{1}\hspace{0ex}d\rho_1
\int_{0}^{1}\hspace{0ex}d\rho_2
\int_{0}^{1}\hspace{0ex}d\rho_3
\,\frac{F(v,E_0\rho_1)}{\rho_1\rho_2}
~,
\end{align}
with
\begin{align}
\vec{\rc} &= 2E_0\sqrt{\zeta_0}\,\rho_1\rho_2\,\mathrm{Rot}_i\Big(\sqrt{1-\zeta_0\rho_2^2}\cos(2\pi\rho_3),\sqrt{1-\zeta_0\rho_2^2}\sin(2\pi\rho_3),-\sqrt{\zeta_0}\,\rho_2\Big)
~.
\end{align}

\subsection{Initial-state terms}
%
\subsubsection{Collinear}
\begin{align}
\LXYZ{\Icolfin}{\inlbl}
&=
\int_0^1d\xP_r\int_0^1d\xM_r
\,\SXYZ{\Icol}{\inlbl}(\xP_r,\xM_r)
\,\theta(\xP_r<1-\xP)
\\&\hspace{4ex}\times
\bigg[
\int\frac{d^{2}p_{r\theperp}}{\pi}\delta_+\big(\sTot\xP_r\xM_r-|p_{r\theperp}|^2\big)
  \ELshiftIin{p_r}\,\theta(\xM_r<1-\xM)
-
\ell_{\inlbl}(\xP+\xP_r)
\bigg]
~,\notag
\end{align}
%
with
\begin{align}
\SXYZ{\Icol}{\inlbl}(\xP_r,\xM_r)
&= 
\theta\big(\xM_r<\xiP\xP_r\big)
\,\frac{-1}{\xM_r(\xP+\xP_r)}
\,\EuScript{Q}_{\inlbl}\bigg(\frac{-\xP_r}{\xP+\xP_r}\bigg)
~.
\end{align}
Writing the $p_{r\theperp}$-integral explicitly, we have
%
\begin{align}
\LXYZ{\Icolfin}{\inlbl}
&=
\int_0^{1}\hspace{0ex}d\xP_r
\int_0^{\xiP\xP_r}\hspace{0ex}d\xM_r
\int_0^1d\rho_3
\,\frac{\tilde{\EuScript{P}}_{\inlbl}(\xP_r/\xP)}{\xP_r\xM_r}
\,\theta(\xP_r<1-\xP)
\notag\\&\hspace{24ex}\times
\Big[
  \ELshiftIin{p_r}\,\theta(\xM_r<1-\xM) - \ell_{\inlbl}(\xP+\xP_r)
\Big]
~,
\end{align}
with
%
\begin{equation}
p_{r\theperp} = \sqrt{\xP_r\xM_r\sTot}\big(\cos(2\pi\rho_3),\sin(2\pi\rho_3)\big)
~,
\end{equation}
%
and where we abbreviate
%
\begin{equation}
\tilde{\EuScript{P}}_{\inlbl}(z)
=
z\,\EuScript{P}_{\inlbl}\bigg(\frac{1}{1+z}\bigg)
~,
\end{equation}
%
which is not singular at $z=0$.
Now, substitute
%
\begin{equation}
\xM_r=\xiP\xP_r\rho_2^2
\quad\textrm{and}\quad
\xP_r=(1-\xP)\rho_1
\end{equation}
%
leading to
%
\begin{align}
\LXYZ{\Icolfin}{\inlbl}
&=
2\int_0^{1}d\rho_1
\int_0^{1}\hspace{0ex}d\rho_2
\int_0^1d\rho_3
\,\frac{\tilde{\EuScript{P}}_{\inlbl}(\xP_r/\xP)}{\rho_1\rho_2}
\notag\\&\hspace{12ex}\times
\Big[
  \ELshiftIin{p_r}\,\theta\big(\xiP(1-\xP)\rho_1\rho_2^2<1-\xM\big)
   - \ell_{\inlbl}(\xP+\xP_r)
\Big]
~,
\end{align}
%
with
%
\begin{equation}
\xP_r=(1-\xP)\rho_1
\;,\;\;
p_{r\theperp} = (1-\xP)\sqrt{\sTot\xiP}\,\rho_1\rho_2\big(\cos(2\pi\rho_3),\sin(2\pi\rho_3)\big)
~.
\end{equation}
%
In the progressive approach this reduces to
%
%
\begin{align}
\LXYZ{\Icolfin}{\inlbl}
&=
2\int_0^{1}d\rho_1
\,\frac{\tilde{\EuScript{P}}_{\inlbl}(\xP_r/\xP)}{\rho_1}\,\ell_{\inlbl}(\xP+\xP_r)
\,\frac{1}{2}\log\bigg(\frac{1-\xM}{\xiP\rho_1(1-\xP)}\bigg)
\,\theta\bigg(\rho_1>\frac{1}{\xiP}\frac{1-\xM}{1-\xP}\bigg)
\notag\\&\hspace{0ex}=
\int_0^{1}dz\,\EuScript{P}_{\inlbl}(z)
\,\frac{\ell(\xP/z)}{z^2}
\,\log\bigg(\frac{1-\xM}{\xiP\xP}\frac{z}{1-z}\bigg)
\,\theta\bigg(\xP<z<\frac{\xiP\xP}{\xiP\xP+1-\xM}\bigg)
~.
\end{align}
%

\subsubsection{Soft}
We need to calculate
\begin{align}
&\LXYZ{\Isoftfin}{\inlbl,b}
=
\frac{-2}{\pi}\int d^{4}p_r\,\delta_+(p_r^2)
\,\frac{1}{E_r^2\,\lop{n_{\inlbl}}{n_r}}
\,F_{\inlbl,b}^{\Isoft}(p_r)
~.
\end{align}
with
%
\begin{align}
F_{\inlbl,b}^{\Isoft}(p_r)
&= A_r^{(\inlbl,b)}\,\theta(x_r<1-x)
 \Big[
  \ELshiftPerp{p_{r}}\theta(\xM_r<1-\xM)\,\theta(E_r<E_0) - \theta\big(E_r<E_0A_r^{(\inlbl,b)}\big)
\Big]
\notag\\&\hspace{0ex}
+
 \Big[
  \theta(1-\xP<E_r/E)\,\theta(E_r<E_0)
 -A_r^{(\inlbl,b)}\,\theta(1-\xP<\xP_r)\,\theta\big(E_r<E_0A_r^{(\inlbl,b)}\big)
\Big] 
\end{align}
%
and
%
\begin{equation}
A_r^{(\inlbl,b)} = \frac{\lop{n_{\inlbl}}{n_b}}{\lop{n_{\inlbl}}{n_r}+\lop{n_b}{n_r}}
~.
\end{equation}
%
We repeat that $F_{\inlbl,b}^{\Isoft}(p_r)\to0$ both when $E_r\to0$ (implying $\xP_r,\xM_r\to0$) and when $n_r\to n_{\inlbl}$ (implying $\xM\to0$, $\xP_r\to E_r/E$, and $A_r^{(\inlbl,b)}\to1$).
Assuming $\pP$ is along the $z$-axis, that is $n_{\inlbl}^\mu=(1,0,0,1)$
\begin{align}
&\LXYZ{\Isoftfin}{\inlbl,b}
=
\frac{-2}{\pi}
  \int_0^{E_0}\hspace{0ex}\frac{dE_r}{E_r}
  \int_{0}^1\hspace{0ex}\frac{d\zeta}{2\zeta}
  \int_0^{2\pi}\hspace{-1ex}d\varphi
 \,F_{\inlbl,b}^{\Isoft}(p_r)
~,
\end{align}
with
\begin{equation}
p_{r,0} = E_r
\;\;,\quad
\vec{p}_r = 2E_r\,\Big(\sqrt{\zeta-\zeta^2}\cos(\varphi),\sqrt{\zeta-\zeta^2}\sin(\varphi),\srac{1}{2}-\zeta\Big)
\;\;,\quad
\lop{n_{\inlbl}}{n_i} = 2\zeta
\;\;.
\end{equation}
%
Substitute $\zeta=\rho_2^2$ and normalize the other variables: 
\begin{align}
&\LXYZ{\Isoftfin}{\inlbl,b}
=
-4
\int_0^1d\rho_1\int_0^1d\rho_2\int_0^1d\rho_3
\,\frac{1}{\rho_1\rho_2}
 \,F_{\inlbl,b}^{\Isoft}(p_r)
~,
\end{align}
with
\begin{equation}
p_{r,0} = \rho_1E_0
\;\;,\quad
\vec{p}_r = 2\rho_1E_0\Big(\rho_2\sqrt{1-\rho_2^2}\cos(2\pi\rho_3),\rho_2\sqrt{1-\rho_2^2}\sin(2\pi\rho_3),\srac{1}{2}-\rho_2^2\Big)
~.
\end{equation}
%

\subsubsection{Soft-collinear}
Using \Equation{Eq:153}, the restriction $\xM_r<\xiP\xP_r$ can be cast in the form of the restriction on the variable $\zeta<\zeta_0$ in the lab frame via
%
\begin{equation}
\xiP = \frac{E}{\bar{E}}\,\frac{\zeta_0}{1-\zeta_0}
\quad\Leftrightarrow\quad
\zeta_0 = \frac{\xiP}{E/\bar{E}+\xiP}
\label{Eq:4419}
~.
\end{equation}
%
Also, we have
%
\begin{equation}
\frac{2}{\sTot}
\,\frac{1}{\xP_r}
\,\frac{1}{\xM_r}
 = \frac{1}{2E\bar{E}}
   \,\frac{E+\bar{E}\xM_r/\xP_r}{E_r}
   \,\frac{2\bar{E}}{E_r\lop{n_{\inlbl}}{n_r}}
= \frac{1}{E_r^2\lop{n_{\inlbl}}{n_r}}
  \,\frac{\bar{E}\xM_r}{E_r-\bar{E}\xM_r}
= \frac{1}{E_r^2\lop{n_{\inlbl}}{n_r}}
  \,\frac{1}{1-\frac{1}{2}\lop{n_{\inlbl}}{n_r}}
~.
\end{equation}
%
We need to calculate
\begin{align}
&\LXYZ{\Isoftcolfin}{\inlbl}
=
\frac{-2C_{\inlbl}}{\pi}\int d^{4}p_r\,\delta_+(p_r^2)
\,\frac{\theta(E_r<E_0)\theta(\lop{n_{\inlbl}}{n_r}<2\zeta_0)}{E_r^2\,\lop{n_{\inlbl}}{n_r}}
\,F_{\inlbl}^{\Isoftcol}(p_r)
~,
\end{align}
with
%
\begin{align}
F_{\inlbl}^{\Isoftcol}(p_r)
&=
  \frac{1}{1-\frac{1}{2}\lop{n_{\inlbl}}{n_r}}
\bigg\{
\theta(\xP_r<1-\xP)\Big[\ELshiftPerp{p_r}\,\theta(\xM_r<1-\xM) - 1\Big]
\notag\\&\hspace{20ex}
+\Big[\theta(1-\xP<E_r/E) - \theta(1-\xP<\xP_r)\Big]
\bigg\}
~.
\end{align}
%
Using that $\pP$ is along the $z$-axis
\begin{align}
&\LXYZ{\Isoftcolfin}{\inlbl}
=
\frac{-2C_{\inlbl}}{\pi}
  \int_0^{E_0}\hspace{0ex}\frac{dE_r}{E_r}
  \int_{0}^{\zeta_0}\hspace{0ex}\frac{d\zeta}{2\zeta}
  \int_0^{2\pi}\hspace{-1ex}d\varphi
\,F_{\inlbl}^{\Isoftcol}(p_r)
~,
\end{align}
with
%
\begin{equation}
p_{r,0} = E_r
\;\;,\quad
\vec{p}_r = 2E_r\,\Big(\sqrt{\zeta-\zeta^2}\cos(\varphi),\sqrt{\zeta-\zeta^2}\sin(\varphi),\srac{1}{2}-\zeta\Big)
\;\;,\quad
\lop{n_{\inlbl}}{n_r} = 2\zeta
\;\;.
\end{equation}
Substitute $\zeta=\eta^2$ and normalize the variables: 
\begin{align}
&\LXYZ{\Isoftcolfin}{\inlbl}
=
-4C_{\inlbl}
\int_0^1d\rho_1\int_0^1d\rho_2\int_0^1d\rho_3
\,\frac{1}{\rho_1\rho_2}
\,F_{\inlbl}^{\Isoftcol}(p_r)
~,
\end{align}
with $p_{r,0}=\rho_1E_0$ and
\begin{equation}
\vec{p}_{r} = 2\rho_1E_0\Big(\rho_2\sqrt{\zeta_0-\zeta_0^2\rho_2^2}\,\cos(2\pi\rho_3),\rho_2\sqrt{\zeta_0-\zeta_0^2\rho_2^2}\,\sin(2\pi\rho_3),\srac{1}{2}-\zeta_0\rho_2^2\Big)
~.
\end{equation}

\end{appendix}

\end{document}